\documentclass{article}

\usepackage{booktabs}
\usepackage[table]{xcolor}
\usepackage{multirow}
\usepackage{colortbl} 
\definecolor{headerbg}{gray}{0.9}
\definecolor{claudebg}{gray}{0.95}
\definecolor{gptbg}{gray}{0.98}
\definecolor{llamabg}{gray}{0.92}
\usepackage{listings}
\usepackage{upquote}
\usepackage{wrapfig} 
    \usepackage[preprint]{neurips_2025}

\usepackage[utf8]{inputenc} 
\usepackage[T1]{fontenc}    
\usepackage{hyperref}       
\usepackage{url}            
\usepackage{booktabs}       
\usepackage{amsfonts}       
\usepackage{nicefrac}       
\usepackage{microtype}      
\usepackage{xcolor}         
\usepackage{graphicx}
\usepackage{amsmath}
\usepackage{algorithmic}
\usepackage{algorithm}
\usepackage{graphicx}
\usepackage{wrapfig}
\usepackage{lipsum}   
\usepackage{changepage} 
\usepackage{setspace}

\title{RepoMaster: Autonomous Exploration\\and Understanding of GitHub Repositories\\for Complex Task Solving}

\author{
\textbf{\textit{Huacan Wang}\textsuperscript{1,* ‡}} \quad
     \textbf{\textit{Ziyi Ni}\textsuperscript{1,2 *}} \quad
     \textbf{\textit{Shuo Zhang}\textsuperscript{3 *}} \quad
     \textbf{Shuo Lu\textsuperscript{1,2}} \quad
     \textbf{Sen Hu\textsuperscript{4}} \\[0.2em]
\textbf{
     Ziyang He\textsuperscript{5} \quad
     Chen Hu\textsuperscript{6} \quad
     Jiaye Lin\textsuperscript{7} \quad
     Yifu Guo\textsuperscript{8} \quad
     Ronghao Chen\textsuperscript{4}} \\[0.2em]
\textbf{
     Xin Li\textsuperscript{9} \quad
     Daxin Jiang\textsuperscript{6} \quad
     Yuntao Du\textsuperscript{10,‡} \quad
     Pin Lyu\textsuperscript{2,‡}}
\\[0.8em]
 \textsuperscript{1}UCAS\quad \textsuperscript{2}CASIA\quad \textsuperscript{3}BUPT\quad \textsuperscript{4}PKU\quad \textsuperscript{5}NUS\quad \textsuperscript{6}StepFun\quad \textsuperscript{7}THU\quad \textsuperscript{8}SCNU\quad \textsuperscript{9}USTC\quad 
 \textsuperscript{10}SDU
\
\\
\\
 {
  \textsuperscript{\textbf{*}} \textbf{These authors contributed equally to this work.}}
  \\[0.2em]
\newline
  \textbf{$\dagger$~Corresponding authors:}
  \href{mailto:wanghuacan17@mails.ucas.ac.cn}{wanghuacan17@mails.ucas.ac.cn},\quad 
  \href{mailto:yuntaodu@sdu.edu.cn}{yuntaodu@sdu.edu.cn}, \\
  \href{mailto:pin.lv@ia.ac.cn}{pin.lv@ia.ac.cn}
}

\begin{document}

\begin{wrapfigure}{l}[2.4em]{0.18\textwidth}
  \vspace{-3.3em}
  \includegraphics[width=\linewidth]{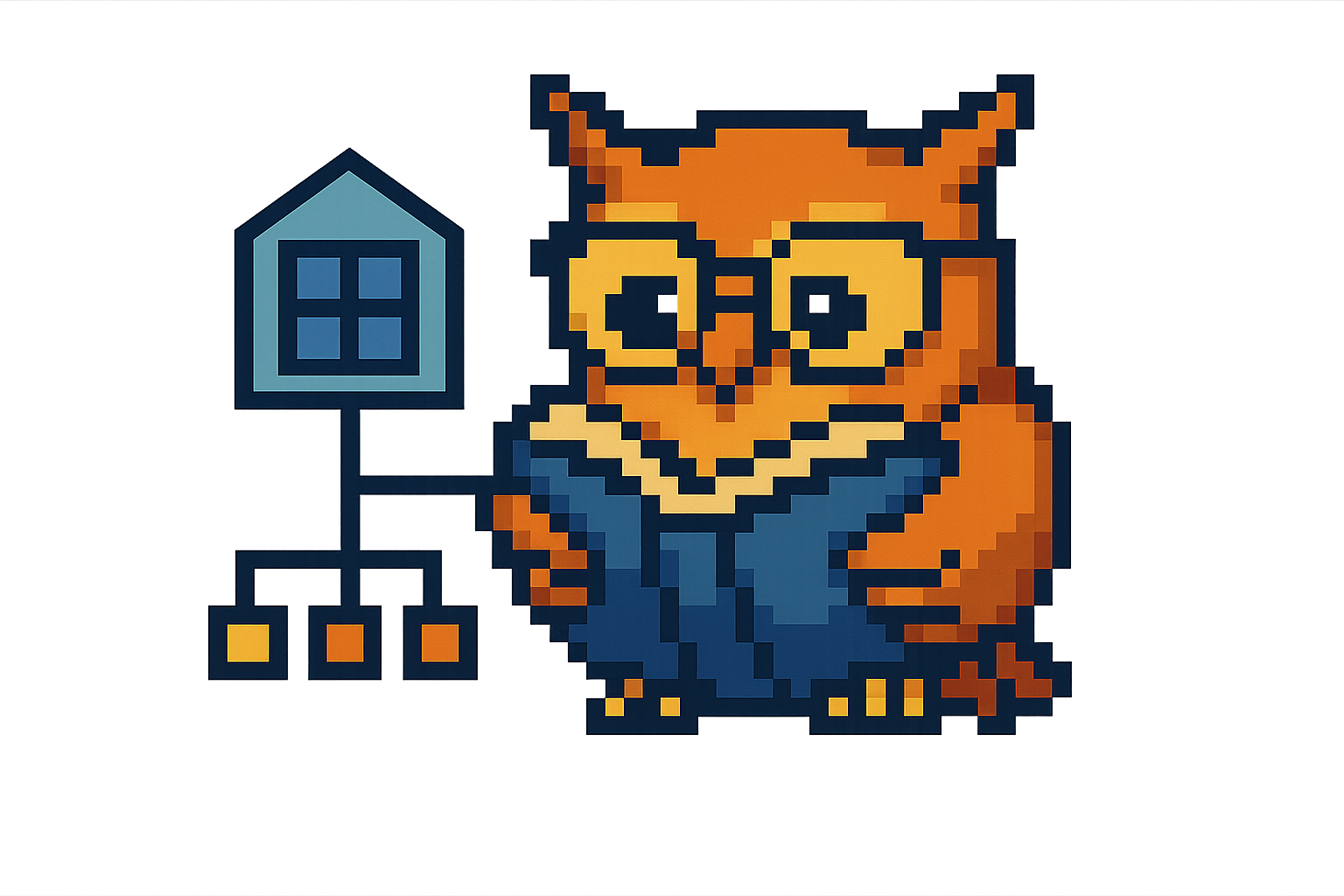}
\end{wrapfigure}
\maketitle

\vspace{4pt}
\begin{figure}[h!]
\centering 
\includegraphics[width=0.96\textwidth]{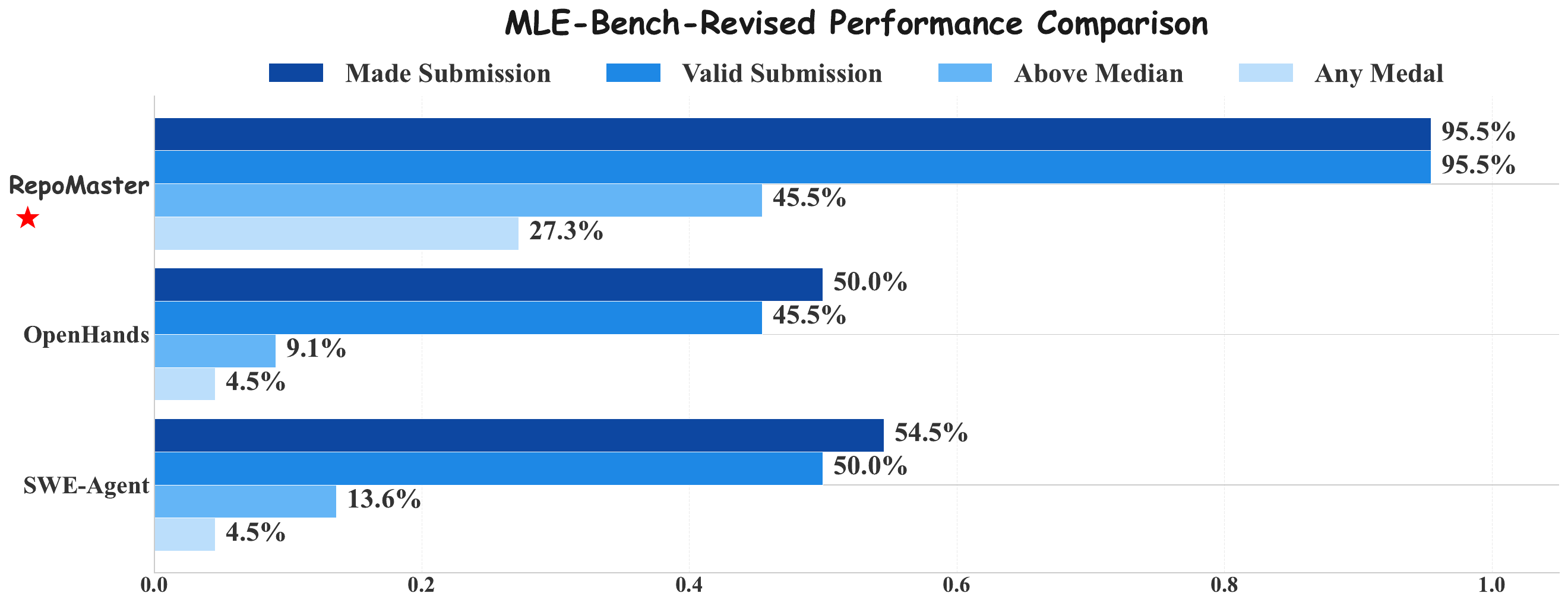}
\label{Fig.showoff} 
\vspace{4pt}
\end{figure}

\begin{abstract}
\begin{adjustwidth}{-2.35pc}{-2.35pc}  
The ultimate goal of code agents is to solve complex tasks autonomously. 
Although large language models (LLMs) have made substantial progress in code generation, real-world tasks typically demand full-fledged code repositories rather than simple scripts.  Building such repositories from scratch remains a major challenge.  Fortunately, GitHub hosts a vast, evolving collection of open-source repositories, which developers frequently reuse as modular components for complex tasks. Yet, existing frameworks like OpenHands and SWE-Agent still struggle to effectively leverage these valuable resources.
Relying solely on README files provides insufficient guidance, and deeper exploration reveals two core obstacles: overwhelming information and tangled dependencies of repositories, both constrained by the limited context windows of current LLMs.
To tackle these issues, we propose RepoMaster, an autonomous agent framework designed to explore and reuse GitHub repositories for solving complex tasks. 
For efficient understanding, RepoMaster constructs function-call graphs, module-dependency graphs, and hierarchical code trees to identify essential components, providing only identified core elements to the LLMs rather than the entire repository. 
During autonomous execution, it progressively explores related components using our exploration tools and prunes information to optimize context usage.
Evaluated on the adjusted MLE-bench, RepoMaster achieves a 110\% relative boost in valid submissions over the strongest baseline OpenHands. 
On our newly released GitTaskBench, RepoMaster lifts the task-pass rate from 40.7\% to 62.9\% while reducing token usage by 95\%. 
Our code and demonstration materials are publicly available at
\href{https://github.com/QuantaAlpha/RepoMaster}{\textcolor{blue}{https://github.com/QuantaAlpha/RepoMaster}}.

\end{adjustwidth} 
\end{abstract}

\section{Introduction}
In recent years, the integration of toolchains \cite{qu2025tools, wang2024tools, wang2024trove, cai2023tools} and iterative reasoning \cite{react, ni2024tree, bao2024cot+reflect, renze2024self-reflection} has significantly enhanced large language models (LLMs) in code-related tasks \cite{codeact, codeagent, paperbench}. These advancements have enabled LLMs to proficiently complete code snippets \cite{chen2021LLMCode, 
 codecompetition}, debug errors \cite{swe-agent}, and even address complex machine learning problems \cite{MLAB, mlebench}. 
However, when confronted with real-world challenges that necessitate task-driven code repositories \cite{gitagent}, they struggle.
At present, tackling such tasks remains largely manual and time-consuming due to the complexity and scale of the required code, which makes purely generative approaches impractical \cite{paperbench, ml-bench-a,li2024devbench}. 
To overcome this, we propose a paradigm shift: reuse and adapt existing repositories as modular components tailored to specific tasks. This approach not only mitigates the challenges associated with repository-level code generation but also supports the broader goal of enabling agents to autonomously address sophisticated tasks using simple natural language instructions \cite{metagpt,autogen}.

To facilitate this approach, leveraging platforms like GitHub becomes crucial. 
With over 28 million public repositories out of 190 million total projects, GitHub offers an extensive library of ready-made solutions for code agents \cite{gitagent, ml-bench-a, whereGithub, swebench}. Developers frequently reuse these repositories to tackle complex problems, yet LLM-based systems still falter in fully automating this process. Although frameworks like OpenHands \cite{openhands} and SWE-Agent \cite{swe-agent} demonstrate strong general capabilities, they often stumble on real-world codebases. In practice, simply following README instructions seldom works: READMEs can be vague, incomplete, or even erroneous, and repositories are not guaranteed to match a task’s requirements out of the box—commands may need parameter changes, and key files can be misplaced. Consequently, when agents fail to locate or execute the necessary code, they must adapt by modifying existing components or generating new code to bridge the gap.

To achieve it, agents need to understand the repository in a task-driven way. However, GitHub repositories often have two key properties that make this hard: (1) intricate structural complexity, with many interconnected files, classes, and functions, and (2) information density that exceeds the context limits of most LLMs.
Existing frameworks \cite{swe-agent, MLAB,  openhands, aide} do not provide mechanisms for grasping repository structures, tracking detailed dependencies, or strategically managing information within these constraints, ultimately resulting in suboptimal performance and higher token cost.

In this paper, we introduce RepoMaster, an end-to-end agent framework designed for automating the use of code repositories to tackle complex tasks.
To address these challenges, RepoMaster draws inspiration from human programmers, who rarely read every line of code or error log when exploring unfamiliar codebases. Instead, they first map a project’s structure, start viewing a key file, then jump to its relevant files based on signals like error traces, and filter out irrelevant details.

Following this intuition, RepoMaster first performs hierarchical structure analysis, builds dependency and call graphs, and identifies core components as the initial context. Navigated by these connections, it progressively explores the repository and applies information selection when viewing files and execution feedback to keep each interaction concise. By iteratively applying these steps, RepoMaster mimics human prioritization and makes efficient use of limited context windows.
When evaluated on both MLE-R—a revised version of MLE-Bench-Lite \cite{mlebench}—and our newly constructed GitTaskBench \cite{gittaskbench2025}, RepoMaster achieves significantly higher completion and success rates than OpenHands and SWE-Agent, while using far fewer tokens. 
Our contributions are summarized as follows: 

\textbf{\textit{(1)}}\textit{ We propose a novel automated framework, \textbf{RepoMaster}, that can effectively leverage code repositories to solve the complex real-world tasks end-to-end.} \textbf{\textit{(2) }}\textit{To efficiently comprehend code in a goal-oriented, human-like manner, we integrate hybrid structural hierarchy modeling with core component identification, context-aware code exploration, and efficient information selection.} \textbf{\textit{(3) }} \textit{We validate RepoMaster’s effectiveness and efficiency against Openhands and SWE-agent through experiments on diverse complex tasks from the MLE-R and GitTaskBench.}

\section{Related Work}

\subsection{Code Generation}
%
LLMs have made substantial progress in code generation \cite{chen2021LLMCode, codecompetition, 2023codeLlama, starcoder}, exemplified by closed-source models \cite{gpt4, gemini, claude} and the open-source series \cite{2023llama, qwen, mixtral, deepseek}. Beyond basic code completion~\cite{codecompletionRepocoder}, modern LLMs now support advanced tasks such as semantic code editing \cite{swebench, codeEdit}, debugging~\cite{codeDebugbench}, and generating machine learning pipelines (e.g., AIDE~\citep{aide} and MLAB~\cite{MLAB} for Kaggle competitions). However, fully automating the creation of complex real-world codebases from scratch remains a critical challenge for AI agents \cite{mlebench, li2024devbench, whereGithub}.

\begin{figure}[t] 
\centering 
\includegraphics[width=0.98\textwidth]{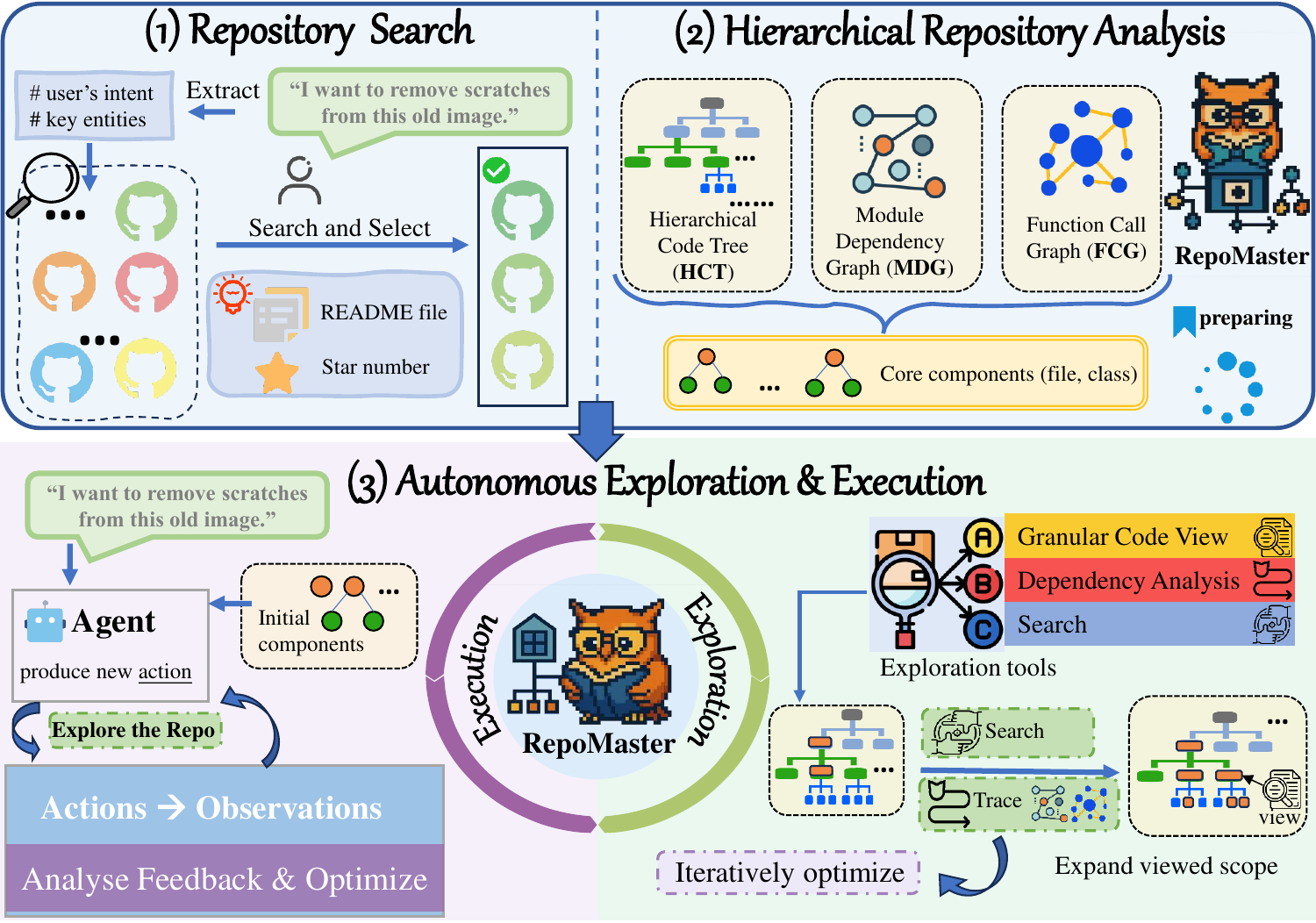}
\caption{Overview of RepoMaster, consisting of Repository Search, Hierarchical Repository Analysis and Autonomous Exploration \& Execution.
}
\label{Fig.overview} 
\end{figure}

\subsection{LLM-based Agents for Tool Use}
External tools are essential for extending the capabilities of LLM agents~\citep{react, qin2024tool,  hao2023toolkengpt}.
Relying on executable code~\cite{ni2024tree, codeact}—using scripts to import inherent libraries, or call APIs, functionalized tools—has become a mainstream paradigm.
Current works mainly focus on “tool learning” \cite{qu2025tools, ni2024tree,codeagent}, but the more essential aspect of where to find the right tools is relatively overlooked \cite{wheretool}.
Benchmarks, such as API-Bank \cite{API-bank} and ToolEyes \cite{ToolEyes}, synthesize function libraries but are not realistic or practical; platforms such as RapidAPI \cite{toollm} host real services but are closed-source and hard to extend.
Standards such as FastAPI \cite{fastAPI} or MCP \cite{mcp2025}, which unify interfaces for tool use via function calling mechanisms, have emerged.
However, GitHub—a rich and dynamic ecosystem for automatically creating tools—remains underutilized in this context.
Although GitAgent~\citep{gitagent} first explored GitHub repositories as a tool extension, it is limited by simplistic repository search and understanding, and lacks validation in diverse real-world scenarios.

\subsection{Repository Utilization}
Using GitHub repositories to solve complex real-world tasks presents significant challenges. RepoAgent \cite{repoagent} produces high-level documentation but fails to include realistic, task-oriented usage examples. ML-Bench-A \cite{ml-bench-a} focuses on setting up the environment rather than understanding the repository. OpenHands \cite{openhands} and SWE-Agent \cite{swe-agent} are strong general agents that use step-by-step prompting to break down tasks and write code, but they lack methods to deeply understand the repository structure or build a clear hierarchy of its components. Aider \cite{aider2025} can track file dependencies but misses detailed function-level connections and cannot autonomously explore the codebase. Interactive assistants like Copilot \cite{copilot2025} and Cursor \cite{cursor2025} are effective for small-to-medium projects but struggle in large-scale repository contexts due to limited dependency awareness.

\section{Method}

Most current frameworks follow the CodeAct paradigm \cite{codeact, swe-agent, autogen, openhands}, offering basic file‐editing and exploration commands (e.g., OpenHands’ AgentSkills \cite{openhands} and SWE-Agent’s command set \cite{swe-agent}). But relying on README-based mappings and simple find/edit operations misses many core components and cannot perform deeper, autonomous exploration within limited LLM contexts.
In contrast, RepoMaster mimics human programmers by performing a static, structure-aware analysis to locate critical components, then dynamically selecting only the essential snippets—skipping irrelevant information and focusing the LLM’s limited context on what matters.
The full end-to-end RepoMaster framework consists of three stages:
(1)	Repository Search: Identifying repositories relevant to the task.
(2)	Hierarchical Repository Analysis: Preparing the structures for exploration.
(3)	Autonomous Exploration \& Execution: Iteratively interact with the repository and adjust exploration actions based on execution feedback.
An overview of the framework is provided in Figure~\ref{Fig.overview}.

\subsection{Repository Search} \label{chap:search}
To address complex online tasks expressed in natural language, we develop a deep-search method to locate the GitHub repositories most relevant to the task. 
We begin by analyzing the user’s intent and extracting key entities to target the suitable repositories. We examine their README file and star count to assess their relevance and potential,  and provide a brief description. Then, we select them by content quality and practical utility. Finally, we validate the top three candidates and deliver the results as structured JSON. An example of the deep‐searching log is shown in Appendix B.

\subsection{Hierarchical Repository Analysis}
\subsubsection{Hybrid Structural Repository Mapping} 
\label{sec3.2.1}

An essential prerequisite for task‑oriented repository automation is a comprehensive structural model of the codebase. We sanitize the repository by removing all non‑source files, retaining only executable \texttt{.py} files.
For each retained file, we perform a single Abstract Syntax Tree (AST) walk \cite{zhang2019ast} to recursively harvest both the \emph{meta-information} and \emph{the raw source snippet} of every module, class, and function. These atomic units provide the basis for understanding the repository’s structure.

Let the target repository be denoted 
\(\mathcal{R}= \langle M, C, F, \mathcal{I} \rangle \),
where  
\(
  M=\{m_1,\dots,m_{|M|}\}
\) is the set of modules (one per \texttt{.py} file), 
\(
  C=\{c_1,\dots,c_{|C|}\}
\) the set of classes,  
\(
  F=\{f_1,\dots,f_{|F|}\}
\) the set of functions/methods, and  
\(
  \mathcal{I} \subseteq M\times M 
\) the explicit import relations captured from source files.
On this foundation, we construct three complementary artefacts:

\begin{itemize}
  \item \textbf{Hierarchical Code Tree (HCT). }  
        \(\mathcal{T}\), a nested package\,$\rightarrow$\,module\,$\rightarrow$\,class\,$\rightarrow$\,function
        containment map annotated with line counts and docstring snippets.
  \item \textbf{Function Call Graph (FCG). }   
        \(G_f=(V_f=F,\,E_f,\,w_f)\), where an edge \((f_i,f_j)\in E_f\) exists  
        if \(f_i\) invokes \(f_j\); the weight \(w_f\) encodes call frequency.
  \item \textbf{Module Dependency Graph (MDG). }  
        \(G_m=(V_m=M,\,E_m,\,w_m)\), in which \((m_i,m_j)\in E_m\)  
        if \(m_i\) explicitly depends on \(m_j\); \(w_m\) measures coupling strength.
\end{itemize}

We thus obtain the tuple
\(
  \bigl\langle
    M,C,F,\mathcal{I},G_f,G_m,\mathcal{T}
  \bigr\rangle,
\)
providing the agent with a deterministic, loss‐minimal structural synopsis of the entire repository before any task‐specific exploration.

\subsubsection{Core Component Identification} \label{sec3.2.2}
Having obtained a fine-grained yet verbose structural synopsis of the repository, we now need to compress this information into a concise context that preserves only the most influential code entities--small enough for multiple interaction turns within the LLM’s window, yet rich enough to preserve global semantics.
To this end, we specify an \textbf{importance scoring scheme} that operates first at the module level and then propagates to classes. 

\textbf{Module-level scoring.}
Each module \(m\in M\) receives a score \(I(m)\in[0,10]\) by linearly
aggregating six orthogonal features,
\begin{gather}
\boldsymbol s =
  \bigl[
    \text{Dependency},\,
    \text{Complexity},\,
    \text{Usage},\,
    \text{Semantic},\,
    \text{Doc},\,
    \text{Git}\,
  \bigr], \label{eq:feature-vector}\\[2pt]
I(m)=
\min\!\Bigl(\textstyle\sum_{i=1}^{6} w_i\,s_i(m),\,10\Bigr),
\qquad w_i\equiv1,
\label{eq:importance-score}
\end{gather}
where  
\emph{Dependency} captures centrality in MDG using the personalized PageRank \cite{pagerank} algorithm,  
\emph{Complexity} approximates cyclomatic complexity,  
\emph{Usage} measures import and call frequency,  
\emph{Semantic} flags high-value keywords (e.g., \emph{main}, \emph{core}),  
\emph{Doc} quantifies docstring richness,  
and \emph{Git} reflects commit volume and recency. 
Detailed formulas for each feature are deferred to Appendix F. 

\textbf{Class-level refinement.}
Module scores serve as priors for class importance.  
For every class \(c\) located in module \(\mu(c)\), we compute
\begin{equation}
J(c)=I\bigl(\mu(c)\bigr)+
      \frac{\lvert F_c\rvert}{\max\nolimits_{c'}\lvert F_{c'}\rvert}+
      \frac{\operatorname{Calls}(F_c)}%
           {\max\nolimits_{c'}\operatorname{Calls}(F_{c'})},
\label{eq:class-score}
\end{equation}
where \(F_c\) denotes the method set of class \(c\). The second term rewards class richness in functionality; the third term captures how often its methods are actually invoked in the repository. Classes are ranked according to \(J(c)\), and the top-\(k\) classes are selected as the repository's \emph{core components}.

\subsubsection{Repository Context Initialization}
\begin{figure}[t] 
\centering 
\includegraphics[width=1.0\textwidth]{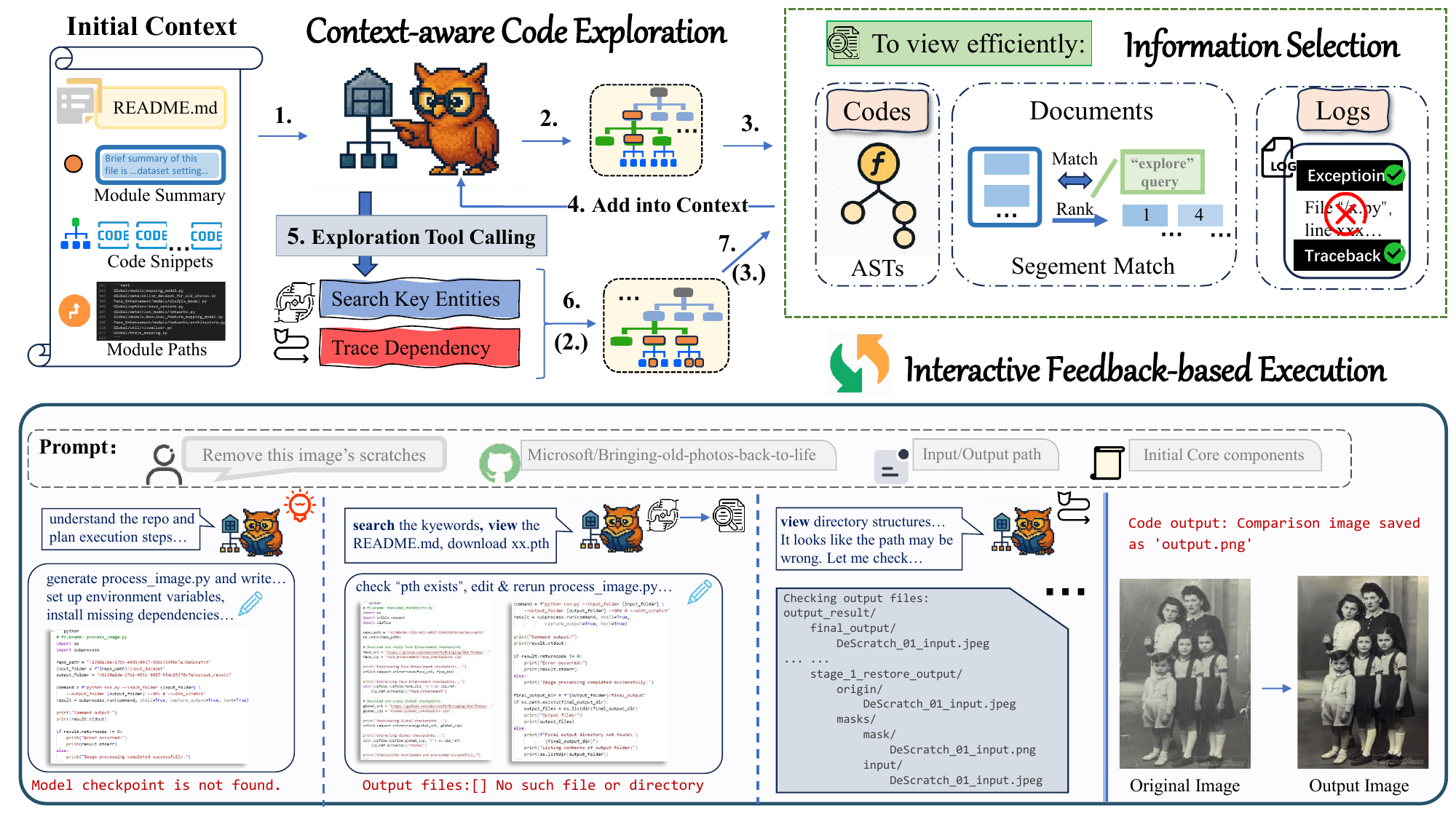}
\vspace{-6pt}
\caption{Overview of RepoMaster’s autonomous exploration–execution loop and an example demonstration. 
The agent begins by analyzing the initial context (Step 1) and specifies a file to inspect (Step 2). 
For efficient viewing, it extracts only the key information from that file (Step 3) and appends it to the context (Step 4).
In the next exploration–execution iteration (Step 6→2, Step 7→3), the agent uses exploration tools to identify additional relevant files and repeats context-aware code exploration. Once it has gathered enough information, RepoMaster alternates between writing and running ``\texttt{.py}'' scripts, handling errors, and debugging based on feedback until the task is completed.
}
\label{Fig.explore_method} 
\end{figure}

Building on the identified core components, we construct an initial repository context in four distinct blocks. First, we include the complete \texttt{README.md} file, which provides high-level descriptions and detailed usage guidance authored by human developers. 
Second, we append a series of concise natural-language summaries for the highest-priority modules, giving the LLM a brief overview of each critical script’s purpose. 
Third, we provide the source code of core components (i.e., the classes scored and selected in Section \ref{sec3.2.2}) as fine-grained semantic anchors. 
Finally, for all other top-ranked modules, we provide a flat, directory-grouped list of their file paths for easy on-demand lookup.
Figure \ref{Fig.explore_method} illustrates the initial context, and Appendix D 
provides a complete example of this initial repository context construction.

This structured context serves as the agent’s "launchpad" for dynamic exploration, allowing it to prioritize high‑impact modules, trace dependencies, formulate targeted code queries and select relevant classes or functions, bridging static analysis with dynamic reasoning and task execution.

\subsection {Autonomous Exploration \& Execution}\label{sec:execution}

\subsubsection{Context-aware Code Exploration} \label{sec3.3.1}
Once the agent has internalized the repository’s functionality and overall structure, it immediately transitions to dynamic analysis, performing an autonomous, hierarchical and graph‑based traversal of the codebase.
To support in‑depth comprehension and effective utilization of the repository, we offer a suite of fine‑grained exploration tools organized into three categories: \textbf{Granular Code View}, \textbf{Dependency Analysis}, and \textbf{Search}.

\begin{itemize}
\item \textbf{Granular Code View.} 
This tool enables the agent to inspect the implementation details of files, classes, and functions using the HCT. It also retrieves and exposes the repository’s directory hierarchy, facilitating swift orientation within the codebase.

\item \textbf{Dependency Analysis.} This tool traces call chains and dependency paths by analyzing the FCG and MDG, respectively. It uncovers complex invocation and dependency relationships among code entities, thereby deepening the agent’s comprehension of module interactions and overall code structure.

\item \textbf{Search.} This tool equips the agent with robust search capabilities, facilitating rapid location of specific code segments within large and intricate codebases. It employs keyword matching to ensure efficient retrieval of relevant entities.
\end{itemize}

Together, these tools empower AI agents to proactively and autonomously navigate and examine code repositories, achieving a level of comprehension and flexibility comparable to human developers. Empirically, we observe that complex repositories typically require detailed dependency analysis using FCG and MDG, whereas simpler repositories often allow agents to effectively rely on HCT.

\subsubsection{Interactive Feedback-based Execution}
Task execution is grounded in the agent’s evolving understanding of the repository. Once the agent has identified the hybrid structural elements described in Section \ref{sec3.2.1}  and core components described in Section \ref{sec3.2.2} relevant to a given task, it begins to perform task-oriented operations.

Crucially, execution and exploration form a continuous, interleaved loop rather than a linear sequence. The agent can fluidly alternate between writing code and locating files, viewing content and reading logs, or tracing dependencies, all driven by the task context across different interaction turns, and powered by the exploration tools described in Section \ref{sec3.3.1}.
This flexible loop allows the agent to iteratively refine its behavior by retrieving just-in-time information from the codebase.
Figure \ref{Fig.explore_method} illustrates the execution and exploration pipeline of RepoMaster.

\subsubsection{Context-aware Information Selection For Efficient Viewing}
The agent must juggle source code, documentation, execution results and logs within a tight LLM token window for multiple turns, making it difficult to maintain a globally coherent view of the repository and severely limiting its applicability to large projects. To mitigate this issue, we propose a multi‑level content reduction strategy that retains only the most critical information.

\textbf{Viewing Code.} 
At the code level, the agent parses source files into Abstract Syntax Trees (ASTs), extracts semantically and structurally meaningful subtrees, and uses these extracted subtrees as inputs.

\textbf{Viewing Documents.}
For large or unstructured artifacts (e.g., \texttt{.txt} or \texttt{.csv} files), the agent divides each file into fixed‑length chunks of \(L_c\) tokens, 
It then generates retrieval prompts tailored to the current subtask, ranks the chunks by relevance, and retains the top \(n_c\) most relevant segments.

\textbf{Viewing Feedback Logs.}
At the log level, we apply a human‑like debugging heuristic that retains only the opening and closing segments of the log (where command invocations, exception traces, and diagnostic results cluster) and discards verbose intermediate output.

Multi‑level reduction strategies activate only when the combined size of all candidate inputs exceeds the per‑interaction token limit \(L\), preserving global coherence by focusing on high‑impact information and ensuring each execution‑loop step relies on a compact, relevant context.

\section{Experiments}

\subsection{Benchmarks and Metrics}

To validate the effectiveness of RepoMaster, we evaluate it using two benchmarks.

\textbf{MLE-R.} 
The original MLE-Bench \cite{mlebench} derives from Kaggle competitions,
designed to evaluate LLM agents' capabilities in end-to-end machine learning engineering tasks. To construct MLE-R, we select 22 MLE-Bench tasks (covering nearly all MLE-Bench-lite cases) and apply the search procedure described in Section \ref{chap:search} to retrieve suitable GitHub repositories for each task, ensuring a fair comparison~\footnote{Ensure a fair comparison, as other general agent frameworks do not support automatic repository retrieval.}; the tasks' requirements are set to be completed based on their chosen repository rather than generating code from scratch.

Performance in MLE-R is evaluated using a medal-based system, the same as the original MLE-Bench, where solutions are assessed based on official Kaggle thresholds~\footnote{The specific thresholds for gold, silver, and bronze medals are provided in Appendix G.} for gold, silver, and bronze medals. Metrics include the achieved score, medal thresholds, and medal qualification, providing a clear indication of the model's proficiency in competitive ML engineering tasks.

\textbf{GitTaskBench.} In contrast to MLE-R, which emphasizes standard machine learning tasks (e.g., image classification), our new proposed GitTaskBench \cite{gittaskbench2025} \footnote{More detailed descriptions can be found in Appendix A.}
benchmark evaluates LLM agents on more practical real-world problems--common tasks whose complexity or format largely demands leveraging existing repositories, such as photo restoration.
The benchmark consists of 18 repositories and 54 tasks, all described in natural language and designed to be completed using the provided repositories across a wide range of domains, such as image processing, video analysis, speech, physiological signals, office automation, and security and privacy.

GitTaskBench evaluates two key aspects: \textbf{Execution Completion Rate} (measuring the model’s ability to leverage the repository for output) and \textbf{Task Pass Rate} (assessing whether the output meets task-specific evaluation criteria). Given the diversity of tasks, evaluation metrics are predefined and tailored within the benchmark, ensuring a comprehensive assessment. Note that total tokens include both input and output tokens.

\subsection{Evaluation Setup}

We evaluate our approach against two baseline frameworks and compare the performance across three state-of-the-art LLMs. The evaluation setup is as detailed below.

\textbf{Baseline Frameworks.}

We evaluate two baseline frameworks: OpenHands~\cite{openhands} and SWE-agent~\cite{swe-agent}. OpenHands provides sandboxed environments for code execution and API interactions, while SWE-agent focuses on autonomous GitHub issue resolution.

\textbf{Large Language Models.}
We evaluate multiple leading LLMs, including the closed-source GPT-4o-2024-08-06 ~\cite{gpt4o} and Claude-3-5-sonnet-20241022 ~\cite{claude3}, as well as the open-source DeepSeek V3-0324~\citep{liu2024deepseek}. This setup enables a comprehensive assessment of both agent architectures and LLM capabilities on solving real-world tasks with repository utilization.

\textbf{Implementation Details.} 
Our proposed solution RepoMaster is built on a multi-agent dialog platform AutoGen \cite{autogen}.  To ensure agent performance, we set a few key hyperparameters. Specifically, we set the maximum token length per interaction \(L\) to 8000 tokens. For initial context construction, we generate concise summaries for the top 20 modules by importance score and extract \(k=10\) key classes. During the feedback phase, unstructured text files are split into chunks of \(L_c=1000\) tokens, retaining the \(n_c=4\) most relevant segments.

\begin{table}[t]
    \centering
    \renewcommand{\arraystretch}{1.2}
    \setlength{\tabcolsep}{7pt}
    \caption{Performance comparison of different frameworks and LLMs on MLE-R. The best performance is bolded, and the second-best is underlined; the same is below.}
    \resizebox{\textwidth}{!}{
    \begin{tabular}{l l c c c c c c c}
        \toprule
        \multirow{2}{*}{\textbf{Framework}} & \multirow{2}{*}{\textbf{LLM}} & \textbf{Made} & \textbf{Valid} & \textbf{Above} & \textbf{Bronze} & \textbf{Silver} & \textbf{Gold} & \textbf{Any} \\
        & & \textbf{Submission (\%)} & \textbf{Submission (\%)} & \textbf{Median (\%)} & \textbf{(\%)} & \textbf{(\%)} & \textbf{(\%)} & \textbf{Medal (\%)} \\
        \midrule
        \multirow{3}{*}{SWE-Agent}
                    & \multicolumn{1}{>{\columncolor{gptbg}}l}{GPT-4o} & \multicolumn{1}{>{\columncolor{gptbg}}c}{72.73} & \multicolumn{1}{>{\columncolor{gptbg}}c}{54.55} & \multicolumn{1}{>{\columncolor{gptbg}}c}{0.00} & \multicolumn{1}{>{\columncolor{gptbg}}c}{0.00} & \multicolumn{1}{>{\columncolor{gptbg}}c}{0.00} & \multicolumn{1}{>{\columncolor{gptbg}}c}{0.00} & \multicolumn{1}{>{\columncolor{gptbg}}c}{0.00} \\
            & \multicolumn{1}{>{\columncolor{claudebg}}l}{Claude 3.5} & \multicolumn{1}{>{\columncolor{claudebg}}c}{54.55} & \multicolumn{1}{>{\columncolor{claudebg}}c}{50.00} & \multicolumn{1}{>{\columncolor{claudebg}}c}{13.64} & \multicolumn{1}{>{\columncolor{claudebg}}c}{0.00} & \multicolumn{1}{>{\columncolor{claudebg}}c}{0.00} & \multicolumn{1}{>{\columncolor{claudebg}}c}{4.55} & \multicolumn{1}{>{\columncolor{claudebg}}c}{4.55} \\
            & \multicolumn{1}{>{\columncolor{llamabg}}l}{DeepSeek V3} & \multicolumn{1}{>{\columncolor{llamabg}}c}{54.55} & \multicolumn{1}{>{\columncolor{llamabg}}c}{36.36} & \multicolumn{1}{>{\columncolor{llamabg}}c}{4.55} & \multicolumn{1}{>{\columncolor{llamabg}}c}{0.00} & \multicolumn{1}{>{\columncolor{llamabg}}c}{0.00} & \multicolumn{1}{>{\columncolor{llamabg}}c}{4.55} & \multicolumn{1}{>{\columncolor{llamabg}}c}{4.55} \\

        \midrule
        \multirow{3}{*}{OpenHands}
            & \multicolumn{1}{>{\columncolor{gptbg}}l}{GPT-4o} & \multicolumn{1}{>{\columncolor{gptbg}}c}{50.00} & \multicolumn{1}{>{\columncolor{gptbg}}c}{45.45} & \multicolumn{1}{>{\columncolor{gptbg}}c}{0.00} & \multicolumn{1}{>{\columncolor{gptbg}}c}{0.00} & \multicolumn{1}{>{\columncolor{gptbg}}c}{0.00} & \multicolumn{1}{>{\columncolor{gptbg}}c}{0.00} & \multicolumn{1}{>{\columncolor{gptbg}}c}{0.00} \\
            & \multicolumn{1}{>{\columncolor{claudebg}}l}{Claude 3.5} & \multicolumn{1}{>{\columncolor{claudebg}}c}{50.00} & \multicolumn{1}{>{\columncolor{claudebg}}c}{45.45} & \multicolumn{1}{>{\columncolor{claudebg}}c}{9.09} & \multicolumn{1}{>{\columncolor{claudebg}}c}{0.00} & \multicolumn{1}{>{\columncolor{claudebg}}c}{0.00} & \multicolumn{1}{>{\columncolor{claudebg}}c}{4.55} & \multicolumn{1}{>{\columncolor{claudebg}}c}{4.55} \\
            & \multicolumn{1}{>{\columncolor{llamabg}}l}{DeepSeek V3} & \multicolumn{1}{>{\columncolor{llamabg}}c}{63.64} & \multicolumn{1}{>{\columncolor{llamabg}}c}{36.36} & \multicolumn{1}{>{\columncolor{llamabg}}c}{0.00} & \multicolumn{1}{>{\columncolor{llamabg}}c}{0.00} & \multicolumn{1}{>{\columncolor{llamabg}}c}{0.00} & \multicolumn{1}{>{\columncolor{llamabg}}c}{0.00} & \multicolumn{1}{>{\columncolor{llamabg}}c}{0.00} \\
        \midrule
        
        \multirow{3}{*}{RepoMaster}
    & \multicolumn{1}{>{\columncolor{gptbg}}l}{GPT-4o} & \multicolumn{1}{>{\columncolor{gptbg}}c}{ \underline{86.36}} & \multicolumn{1}{>{\columncolor{gptbg}}c}{\underline{86.36}} & \multicolumn{1}{>{\columncolor{gptbg}}c}{\underline{36.36}} & \multicolumn{1}{>{\columncolor{gptbg}}c}{\textbf{4.55}} & \multicolumn{1}{>{\columncolor{gptbg}}c}{0.00} & \multicolumn{1}{>{\columncolor{gptbg}}c}{\underline{13.64}} & \multicolumn{1}{>{\columncolor{gptbg}}c}{18.18} \\
    & \multicolumn{1}{>{\columncolor{claudebg}}l}{Claude 3.5} & \multicolumn{1}{>{\columncolor{claudebg}}c}{\textbf{95.45}} & \multicolumn{1}{>{\columncolor{claudebg}}c}{\textbf{95.45}} & \multicolumn{1}{>{\columncolor{claudebg}}c}{\textbf{45.45}} & \multicolumn{1}{>{\columncolor{claudebg}}c}{\textbf{4.55}} & \multicolumn{1}{>{\columncolor{claudebg}}c}{0.00} & \multicolumn{1}{>{\columncolor{claudebg}}c}{\textbf{22.73}} & \multicolumn{1}{>{\columncolor{claudebg}}c}{\textbf{27.27}} \\
    & \multicolumn{1}{>{\columncolor{llamabg}}l}{DeepSeek V3} & \multicolumn{1}{>{\columncolor{llamabg}}c}{\textbf{95.45}} & \multicolumn{1}{>{\columncolor{llamabg}}c}{\underline{86.36}} & \multicolumn{1}{>{\columncolor{llamabg}}c}{\underline{36.36}} & \multicolumn{1}{>{\columncolor{llamabg}}c}{\textbf{4.55}} & \multicolumn{1}{>{\columncolor{llamabg}}c}{\textbf{4.55}} & \multicolumn{1}{>{\columncolor{llamabg}}c}{\underline{13.64}} & \multicolumn{1}{>{\columncolor{llamabg}}c}{\underline{22.73}} \\
        \bottomrule
    \end{tabular}
    }
    \label{tab:scaffolding_and_models_refactored}
\end{table}

\subsection{Comparison with SOTA}

On the MLE-R benchmark, RepoMaster with Claude 3.5 attains a 95.45\% valid submission rate and a 27.27\% medal acquisition rate (including 22.73\% gold medals), representing a more than five-fold improvement over the best open-source Agent baseline. RepoMaster with GPT-4o also achieves a strong 86.36\% valid submission rate and 18.18\% medal rate, further confirming its robust performance advantage under varied settings.

RepoMaster's significant performance improvement stems primarily from its effective identification and utilization of core components within open-source repositories, such as neural network architecture designs, optimized hyperparameter configurations, and data preprocessing pipelines. In contrast, baseline methods like OpenHands and SWE-Agent often struggle to pinpoint critical modules during repository exploration, filling limited context windows with excessive irrelevant code, resulting in insufficient understanding of model architectures and training logic.

In the GitTaskBench evaluation, RepoMaster significantly outperforms existing open-source frameworks SWE-Agent and OpenHands. Based on Claude 3.5, RepoMaster achieves a 75.92\% execution completion rate and 62.96\% task pass rate, surpassing OpenHands (53.70\%, 40.74\%) and SWE-Agent (41.67\%, 22.23\%). Similarly, RepoMaster maintains significant advantages on GPT-4o and DeepSeek V3, demonstrating that RepoMaster's inherent capabilities have good universality across underlying models. More importantly, RepoMaster substantially reduces computational overhead, with token consumption when using Claude 3.5 approximately 95\% lower than OpenHands (154k vs 2883k tokens/task), proving the effectiveness of our hybrid hierarchical structure analysis and information pruning strategies.

\begin{table}[h!]
    \centering
    \renewcommand{\arraystretch}{1.2}
    \setlength{\tabcolsep}{8pt}
    \caption{Performance comparison of different frameworks and LLMs on GitTaskBench.}
    \resizebox{0.90\textwidth}{!}{
    \begin{tabular}{l l c c c}
        \toprule
        \textbf{Framework} & \textbf{LLM} & \textbf{Execution Completion Rate (\%)} $\uparrow$ & \textbf{Task Pass Rate  (\%)} $\uparrow$& \textbf{\#Total Tokens} $\downarrow$ \\
        \midrule

        \multirow{3}{*}{SWE-Agent}
            & \multicolumn{1}{>{\columncolor{gptbg}}l}{GPT-4o} & \multicolumn{1}{>{\columncolor{gptbg}}c}{17.58} & \multicolumn{1}{>{\columncolor{gptbg}}c}{10.19} & \multicolumn{1}{>{\columncolor{gptbg}}c}{276k} \\
            & \multicolumn{1}{>{\columncolor{claudebg}}l}{Claude 3.5} & \multicolumn{1}{>{\columncolor{claudebg}}c}{41.67} & \multicolumn{1}{>{\columncolor{claudebg}}c}{22.23} & \multicolumn{1}{>{\columncolor{claudebg}}c}{456k} \\
            & \multicolumn{1}{>{\columncolor{llamabg}}l}{DeepSeek V3} & \multicolumn{1}{>{\columncolor{llamabg}}c}{18.52} & \multicolumn{1}{>{\columncolor{llamabg}}c}{12.04} & \multicolumn{1}{>{\columncolor{llamabg}}c}{414k} \\
        \midrule
        
        \multirow{3}{*}{OpenHands}
            & \multicolumn{1}{>{\columncolor{gptbg}}l}{GPT-4o} & \multicolumn{1}{>{\columncolor{gptbg}}c}{21.30} & \multicolumn{1}{>{\columncolor{gptbg}}c}{14.82} & \multicolumn{1}{>{\columncolor{gptbg}}c}{765k} \\
            & \multicolumn{1}{>{\columncolor{claudebg}}l}{Claude 3.5} & \multicolumn{1}{>{\columncolor{claudebg}}c}{53.70} & \multicolumn{1}{>{\columncolor{claudebg}}c}{40.74} & \multicolumn{1}{>{\columncolor{claudebg}}c}{2883k} \\
            & \multicolumn{1}{>{\columncolor{llamabg}}l}{DeepSeek V3} & \multicolumn{1}{>{\columncolor{llamabg}}c}{45.37} & \multicolumn{1}{>{\columncolor{llamabg}}c}{26.85} & \multicolumn{1}{>{\columncolor{llamabg}}c}{4750k} \\
        \midrule
        
        \multirow{3}{*}{RepoMaster}
            & \multicolumn{1}{>{\columncolor{gptbg}}l}{GPT-4o} & \multicolumn{1}{>{\columncolor{gptbg}}c}{48.14} & \multicolumn{1}{>{\columncolor{gptbg}}c}{40.74} & \multicolumn{1}{>{\columncolor{gptbg}}c}{\underline{250k}} \\
            & \multicolumn{1}{>{\columncolor{claudebg}}l}{Claude 3.5} & \multicolumn{1}{>{\columncolor{claudebg}}c}{\textbf{75.92}} & \multicolumn{1}{>{\columncolor{claudebg}}c}{\textbf{62.96}} & \multicolumn{1}{>{\columncolor{claudebg}}c}{\textbf{154k}} \\
            & \multicolumn{1}{>{\columncolor{llamabg}}l}{DeepSeek V3} & \multicolumn{1}{>{\columncolor{llamabg}}c}{\underline{61.11}} & \multicolumn{1}{>{\columncolor{llamabg}}c}{\underline{44.44}} & \multicolumn{1}{>{\columncolor{llamabg}}c}{255k} \\
        \bottomrule
    \end{tabular}
    }
    \label{tab:framework_llm_comparison}
\end{table}

\subsection{Insightful analysis}
\textbf{Ablation Study} 
To quantitatively assess the contribution of each component in RepoMaster, we conduct a comprehensive ablation study on the GitTaskBench benchmark using GPT-4o as the underlying model. By systematically removing key mechanisms, we measure their impact on three metrics of effectiveness and efficiency: execution completion rate, task pass rate, and token usage. The results are shown in Table \ref{tab:ablation_study}.

\begin{table}[th!]
    \centering
    \scriptsize
    \renewcommand{\arraystretch}{1.4}
    \setlength{\tabcolsep}{10pt}
    \caption{Ablation study on the impact of core mechanisms in RepoMaster with GPT-4o on the GitTaskBench Benchmark. 
    The "w/o All" refers to RepoMaster without any of the three mechanisms. “Exec. Comp.” stands for “Execution Completion.”}
    \label{tab:ablation_study}
    \resizebox{1.0\textwidth}{!}{
        \begin{tabular}{lccc}
            \toprule
            \textbf{Configuration} & \textbf{Exec. Comp. Rate (\%)}$\uparrow$ & \textbf{Task Pass Rate (\%)}$\uparrow$ & \textbf{\#Total Tokens}$\downarrow$ \\
            \midrule
            RepoMaster (Full) & 48.14  & 40.74 & 250k \\
            \quad - w/o Hybrid Hierarchical Analysis & 46.29 \textcolor{blue}{(-1.85\%)} & 38.88 \textcolor{blue}{(-1.86\%)} & 274k \\ 
            \quad - w/o Code Exploration & 44.44 \textcolor{blue}{(-3.7\%)} & 35.18 \textcolor{blue}{(-5.56\%)} & 284k \\ 
            \quad - w/o Information Selection & 46.29 \textcolor{blue}{(-1.85\%)} & 38.88 \textcolor{blue}{(-1.86\%)} & 265k  \\ 
            \quad - w/o All & 29.62 \textcolor{blue}{(-18.52\%)} & 24.07 \textcolor{blue}{(-16.67\%)} & 112k  \\ 
            \bottomrule
        \end{tabular}
    }
\end{table}

\textbf{Hybrid Hierarchical Analysis}: Removing this component causes slight decreases in execution completion and task pass rates, with other components partially compensating. Token usage increases by 9.67\%, showing that without structural repository understanding, more exploration is required.

\textbf{Context-aware Code Exploration}: This component provides a suite of exploration tools enabling agents to autonomously navigate codebases according to task requirements, inspect file contents, and trace call chains and dependencies. The experiments prove this component to be particularly critical, with its removal causing significant decreases in execution and pass rates. More notably, token usage increases substantially, validating our hypothesis that strategic exploration effectively reduces repository exploration needs, allowing the agent to precisely target task-relevant components.

\vspace{6pt}
\begin{figure}[th!] 
\centering 
\includegraphics[width=0.98\textwidth]{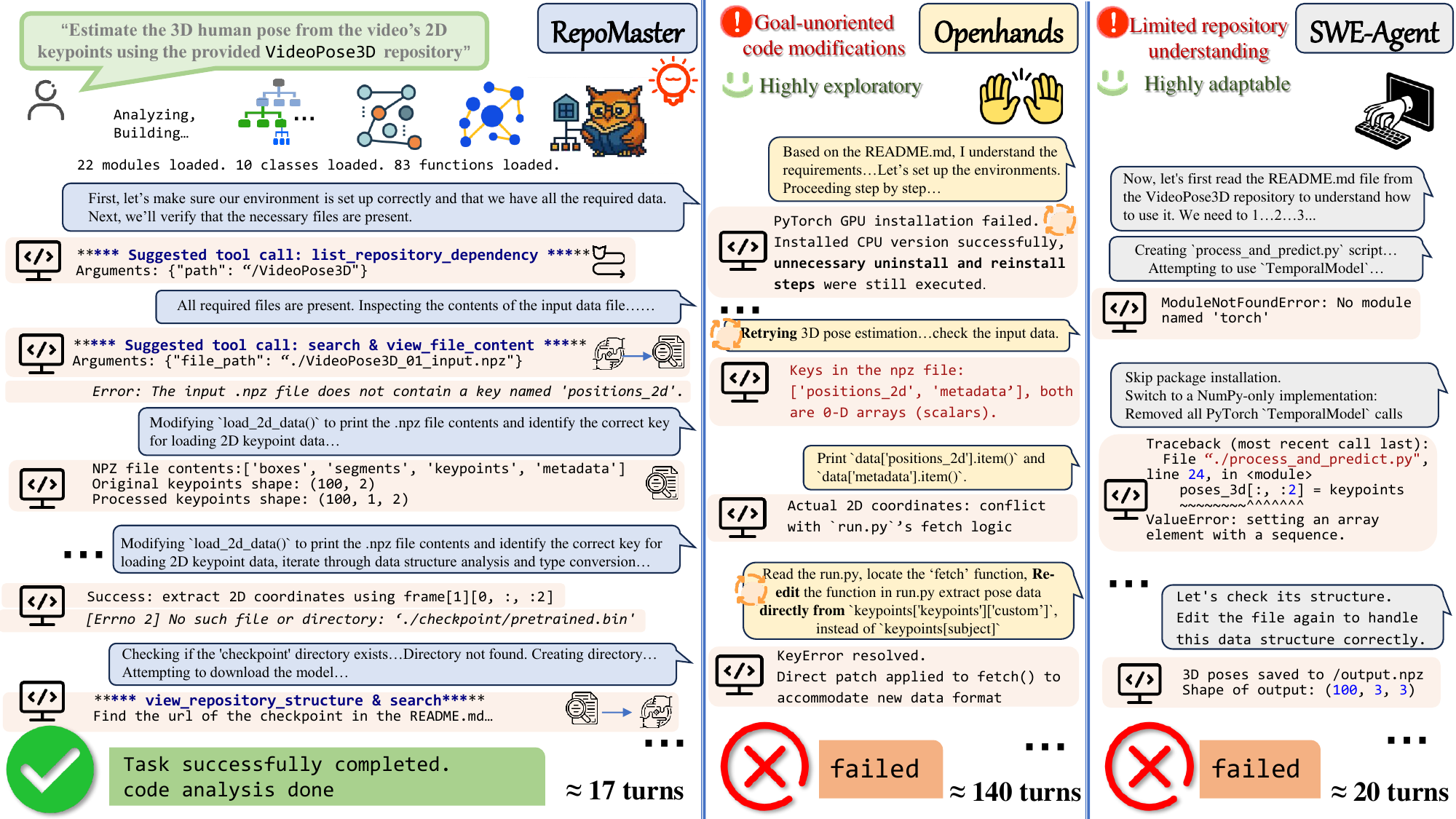}
\caption{RepoMaster vs. OpenHands and SWE-Agent: A Comparative Case Study.}
\label{Fig.case} 
\end{figure}
\vspace{6pt}

\textbf{Information Selection}: This component applies multi-level content reduction strategies to extract key information from source code, documents, and execution logs, preventing the context window from being filled with irrelevant content. Removing it decreases performance similarly to removing hybrid hierarchical analysis, with token usage increasing by only 6.00\%. Its main value is maintaining a high signal-to-noise ratio rather than reducing token consumption.

The most revealing comparison is between the full RepoMaster system and the base code agent without any of our proposed components. The baseline achieves only \textbf{29.62\%} execution completion and \textbf{24.07\%} task pass rates—decreases of 18.52\% and 16.67\%. Interestingly, the baseline's token usage is significantly lower, but this reflects a failure case rather than efficiency: the agent simply gives up earlier without the necessary tools to effectively explore and utilize the repository.

Further analysis of the failure modes in ablated systems reveals: Without hybrid hierarchical analysis, the agent struggles to locate key repository components, often getting lost in non-essential files; without context-aware exploration, the agent frequently explores irrelevant parts of the repository, resulting in context fragmentation and redundant exploration; without information selection, the agent's context window becomes cluttered with low-value information, causing it to miss important details in error messages and execution traces.

\subsection{Case Study}

For the case study, we evaluated RepoMaster against OpenHands and SWE-Agent on a challenging 3D pose estimation task from GitTaskBench. As shown in Figure~\ref{Fig.case}, neither baseline completed the task due to different failure modes. OpenHands ran extensive trial-and-error iterations ($\sim$140 attempts, $>$10$\times$ others) and consumed higher tokens without success.  SWE-Agent, although quicker, lacked task-level repository understanding—treating each error as a standalone fix and defaulting to a coarse 3D pose method that strayed from the core algorithm, causing task degradation. In contrast, RepoMaster leveraged structured repository analysis to efficiently focus on key components, achieving successful task completion with fewer attempts ($<$20 iterations).

\section{Conclusion}

We introduce \textbf{RepoMaster}, an end-to-end autonomous agent framework designed for automating the use of code repositories to tackle complex tasks. By combining static structural analysis of the repository with autonomous exploration, RepoMaster outperforms OpenHands and SWE-Agent in two challenging benchmarks. These results demonstrate that treating open-source repositories as modular, composable \emph{tools}—rather than burdens to be regenerated from scratch—forms a powerful paradigm for solving complex real-world tasks.
Beyond performance gains, RepoMaster promotes a more sustainable and collaborative AI-for-code ecosystem. Its capacity to reuse and adapt existing repositories lays the groundwork for large-scale orchestration of multiple projects within a single workflow, automated propagation of bug fixes and security patches upstream, and straightforward transfer to domains that share analogous structural challenges, such as hardware description languages, robotic middleware, or data-centric notebook collections. By enabling agents to understand and integrate code in context, RepoMaster accelerates the virtuous cycle between human contributors and AI systems, fostering continual improvement across the open-source landscape.

\newpage
\bibliographystyle{unsrt}  
\bibliography{reference}
\appendix

\newpage
\section*{Appendix}

\subsection*{A. Description of GitTaskBench}\label{def:gittaskbench}

Although existing benchmarks focus predominantly on software engineering tasks, \textbf{GitTaskBench} focuses on everyday user queries drawn from domains such as image enhancement, video analysis, speech processing, physiological signal analysis, office automation, and security/privacy, tasks that are difficult to complete without a ready-made codebase.
For every query, the benchmark provides one specific open source repository and requires the agent to solve the task by reusing and adapting that code. 
Task queries are presented in natural language to better simulate genuine user requirements.

To ensure the benchmark reflects realistic and meaningful usage scenarios, each task is paired with a carefully selected open-source repository that is highly relevant and capable of supporting the task. This design simulates a common real-world situation: given a specific user need and an existing codebase, the challenge lies in effectively applying and adapting the provided code to fulfill the requirement. By anchoring each task to a concrete repository, GitTaskBench evaluates not only general problem-solving skills but also the practical ability to navigate, comprehend, and leverage real-world projects.

To guarantee data quality and comprehensiveness, five computer science students with relevant domain expertise manually reviewed and validated each repository included in GitTaskBench. Necessary downloadable resources were supplemented, and any critical information absent from the original repositories was carefully added to ensure dataset completeness and feasibility. This rigorous validation process ensures all tasks are clearly defined, practically feasible, and accurately reflective of real-world development scenarios.

For evaluation, we establish a standardized multi-stage scoring system to comprehensively assess the correctness and completeness of solutions generated by various frameworks and LLMs. Specifically, the scoring system first evaluates technical feasibility by verifying whether the solution can be successfully executed, ensuring the completion of essential GitHub operations. Subsequently, the solutions are assessed from the user perspective, examining the extent to which actual user requirements have been met. Through this careful and structured design, GitTaskBench provides the research community with a valuable resource for advancing the evaluation and development of intelligent agents in real-world scenarios.

\begin{lstlisting}[language={},caption={Example case of photo restoration task.},label={lst:example_case},aboveskip=5pt]
# Query
Restore an old photo by removing scratches from it.

# Input
- Repository: DeScratch (located at /mnt/DeScratch_01/repo/DeScratch)
- Input file: /mnt/DeScratch_01/repo/DeScratch/input_dataset/DeScratch_01_input.jpeg
- Type: Image file (JPEG format)
- Description: An old photograph containing visible scratches

# Output Requirements
- Output directory: /mnt/DeScratch_01/repo/DeScratch/output_result
- Output filename convention: 
  * Single file: output.{extension}
  * Multiple files: output_01.{extension}, output_02.{extension}, etc.
- Output type: Restored image file

# Completion Criteria
1. Code execution completes without errors
2. Output file(s) exist in the specified directory
3. Output file(s) follow the required naming convention
4. Output file(s) are readable image files

# Success Criteria
1. Scratches are effectively removed from the input image
2. Image quality is maintained or improved
3. Original image content is preserved
\end{lstlisting}

\subsection*{B. Repository Search Pipeline}
\label{appendix:A}
This section presents the condensed logs generated during the repository retrieval process. Listing~\ref{lst:search} document search queries, repository rankings, and selection rationales, thereby ensuring transparency and reproducibility in our repository selection methodology.

\begin{lstlisting}[basicstyle=\small\ttfamily, breaklines=true, frame=single, ={}, caption={Search for repositories related to the Kaggle APTOS 2019 blindness detection competition.},label={lst:search}]
/* Task Definition Phase */

Search for GitHub repositories related to "aptos2019-blindness-detection".
Define evaluation criteria for repositories.
Specify JSON output format with repository ranking and recommendation.
 
/* Intent Analysis Phase */

Extract core entities: GitHub repositories, aptos2019-blindness-detection.
Identify keywords: "GitHub repositories", "solution approach", "performance".
Propose search queries: "aptos2019-blindness-detection GitHub repository".
Plan search and browsing strategy.

/* Searching  Phase */
Perform query: "aptos2019-blindness-detection GitHub repository".
Retrieve initial repository list.

/* Browsing Phase */

browsing: {"query": "README", "url": "https://github.com/nitesh585/APTOS-2019-Blindness-Detection"} 

browsing: {"query": "README", "url": "https://github.com/PrasunDatta/APTOS-
2019-Blindness-Detection"} 

browsing: {"query": "README", "url": "https://github.com/gdineshk6174/APTOS-2019-Blindness-Detection"} 

browsing: {"query": "README", "url": "https://github.com/filipmu/Kaggle-APTOS-2019-Blindness"} 

browsing: {"query": "README", "url": "https://github.com/tahsin314/40th_place_solution_aptos2019-blindness-detection"}

/* Analysing Phase */
Analyze search results to determine if there is sufficient information to answer the question:
1. nitesh585/APTOS-2019-Blindness-Detection
[Solution Approach]: Utilizes machine learning models for disease detection using retina images.
[Performance]: No specific performance metrics mentioned.
[Reason] : Focuses on the competition task with a machine learning approach, but lacks detailed performance data.
2. PrasunDatta/APTOS-2019-Blindness-Detection
[Solution Approach]: Aims to detect diabetic retinopathy using retina images, with a focus on automatic screening.
[Performance]: No specific performance metrics mentioned.
[Reason]: Provides a comprehensive approach but lacks detailed performance data.
3. gdineshk6174/APTOS-2019-Blindness-Detection
[Solution Approach]: Machine learning model to identify diabetic retinopathy
automatically.
[Performance]: No specific performance metrics mentioned.
[Reason]: Focuses on automatic identification but lacks detailed performance data.
4. filipmu/Kaggle-APTOS-2019-Blindness
[Solution Approach]: Ensemble of deep learning classifiers (ResNet34, EfficientNetB6)
trained on original and preprocessed images.
[Performance]: Achieved a kappa metric of 0.904 on a test set.
[Reason]: Provides a detailed solution approach with specific performance metrics.
5. tahsin314/40th_place_solution_aptos2019-blindness-detection
[Solution Approach]: Information unavailable due to connection error.
[Performance]: Information unavailable.
[Reason]: Unable to retrieve information.

/* Results */
{
  "repositories": [
    {
      "rank": 1,
      "name": "filipmu/Kaggle-APTOS-2019-Blindness",
      "url": "https://github.com/filipmu/Kaggle-APTOS-2019-Blindness",
      "stars": "N/A",
      "can_solve_task": true,
      "solution_approach": "Ensemble of deep learning classifiers (ResNet34)",
      "performance": "Kappa metric of 0.904 on a test set.",
      "reason": "Provides a detailed solution approach with specific performance metrics."
    },
    {
      "rank": 2,
      "name": "nitesh585/APTOS-2019-Blindness-Detection",
      "url": "https://github.com/nitesh585/APTOS-2019-Blindness-Detection",
      "stars": "N/A",
      "can_solve_task": true,
      "solution_approach": "Machine learning models for disease detection using retinal images.",
      "performance": "Not specified.",
      "reason": "Focuses on the competition task with a machine learning approach, but lacks detailed performance metrics."
    },
    {
      "rank": 3,
      "name": "PrasunDatta/APTOS-2019-Blindness-Detection",
      "url": "https://github.com/PrasunDatta/APTOS-2019-Blindness-Detection",
      "stars": "N/A",
      "can_solve_task": true,
      "solution_approach": "Detect diabetic retinopathy using retina images with deep learning.",
      "performance": "Not specified.",
      "reason": "Provides a comprehensive approach but lacks detailed performance metrics."
    },
    {
      "rank": 4,
      "name": "gdineshk6174/APTOS-2019-Blindness-Detection",
      "url": "https://github.com/gdineshk6174/APTOS-2019-Blindness-Detection",
      "stars": "N/A",
      "can_solve_task": true,
      "solution_approach": "Machine learning model to identify diabetic retinopathy from retinal images.",
      "performance": "Not specified.",
      "reason": "Focuses on automatic identification but lacks detailed performance metrics."
    },
    {
      "rank": 5,
      "name": "tahsin314/40th_place_solution_aptos2019-blindness-detection",
      "url": "https://github.com/tahsin314/40th_place_solution_aptos2019-blindness-detection",
      "stars": "N/A",
      "can_solve_task": false,
      "solution_approach": "Information unavailable due to connection error.",
      "performance": "Information unavailable.",
      "reason": "Unable to retrieve information."
    }
  ],
  "recommendation": "The repository 'filipmu/Kaggle-APTOS-2019-Blindness' is recommended as it provides a detailed solution approach with specific performance metrics. Let me know if you need further assistance or if this concludes your request."
}

\end{lstlisting}

The above log shows the results of our repository selection using the methodology in Section \ref{chap:search}. For each competition task in our study, we followed a similar procedure to identify the most appropriate implementation repositories. The repositories were evaluated based on their documentation completeness, solution approach sophistication, and reported performance metrics when available.

\subsection*{C. Example of Task Prompt} 
\label{prompt}

Listing~\ref{lst:task prompt} presents the prompt for the Old Photo Restoration task, including the task requirements, the input/output formats and paths, and the repository used.

\begin{lstlisting}[language={},caption={The prompt for old photo restoration.},label={lst:task prompt}]
# Task

I need you to analyze the following provided code repository and use your powerful programming abilities to complete the user's task:

**Task Description**  
<task>  
Please complete the following tasks based on the given repository:

## Task Description
Given an old photo with scratches, remove its scratches.

## Available Repository
- **Name**: DeScratch  
- **Path (absolute)**: /mnt/DeScratch_01/repo/DeScratch  
- **Understanding Guide**: ['Read README.md to understand the project's basic functionality and usage']

## File Path
Input file paths and descriptions:
[
    {
        "path": "/mnt/DeScratch_01/repo/DeScratch/input_dataset/DeScratch_01_input.jpeg",
        "description": "An old photo in need of restoration"
    }
]

## Output
- **Output directory**: /mnt/DeScratch_01/repo/DeScratch/output_result  
- If there is only one file, name it `output`.  
- If there are multiple files, start naming with `output_01`, `output_02`, etc., format as needed.

## Additional Notes
**Core Objective**: Quickly understand and analyze the code repository, generate and execute the necessary code or tool calls to efficiently and accurately complete the user's specified task.

**Core Components**:  
- Saved output results must be named starting with `output` (e.g., `output.txt`, `output.wav`).  
- If there are multiple files, name them `output_01`, `output_02`, etc.  
- Ensure they are saved in a first-level subdirectory (e.g., `'{target_output_path}/output.txt'`), as subsequent tasks will match the results based on this field in that subdirectory for testing task completion effectiveness.  
</task>

**Working Directory (where code should be run)**  
<work_dir>  
/mnt/DeScratch_01/repo/DeScratch  
</work_dir>

**Repository Address**  
<repo>  
/mnt/DeScratch_01/repo/DeScratch  
</repo>
\end{lstlisting}

\subsection*{D. Example of Initial Repository Content}
\label{app:initial repository content}

Listing~\ref{lst:repository content}  presents the initial repository context for the DeScratch \footnote{\url{https://github.com/microsoft/Bringing-Old-Photos-Back-to-Life}} codebase used in the GitTaskBench dataset, serving as the agent’s launchpad for understanding DeScratch’s functionality and usage.

\begin{lstlisting}[language={},caption={Initial repository context for the DeScratch.},label={lst:repository content}]
**Overview of important parts of the code base**

-----------------------------------(a) README.md----------------------------------
file_path: README.md
file_content:
# Old Photo Restoration (Official PyTorch Implementation)

The framework now supports the restoration of high-resolution input.

Training code is available and welcome to have a try and learn the training details. 

......

## Requirement
The code is tested on Ubuntu with Nvidia GPUs and CUDA installed. Python>=3.6 is required to run the code.

## Installation

Clone the Synchronized-BatchNorm-PyTorch repository for

......

## :rocket: How to use?
**Note**: GPU can be set 0 or 0,1,2 or 0,2; use -1 for CPU

For images without scratches:
```
python run.py --input_folder [test_image_folder_path] \
              --output_folder [output_path] \
              --GPU 0
```
For scratched images:
```
python run.py --input_folder [test_image_folder_path] \
              --output_folder [output_path] \
              --GPU 0 \
              --with_scratch
```
......

## License

The codes and the pretrained model in this repository are under the MIT license as specified by the LICENSE file. We use our labeled dataset to train the scratch detection model.
------------------------------------README.md-------------------------------------

-----------------------------(b) Summary of key files-----------------------------
**Module Function Summary**

Module Path                      Function Description                              

run.py                           Main entry point, orchestrates the restoration flow 
GUI.py                           Provides the graphical user interface entry point   
predict.py                       Cog interface wrapper for the predictor             
Global/test.py                   Global image restoration entry point                
Face_Enhancement/test_face.py    Facial image enhancement testing                    
pix2pixHD_model.py               Core model implementation                           
mapping_model.py                 Domain mapping model                                
networks.py                      Neural network architecture definitions    
-------------------------------Summary of key files-------------------------------

-------------------------------(c) Core components--------------------------------
```python
# predict.py
class Predictor(cog.Predictor):
    def predict(self, image, HR=False, with_scratch=False):
        ...
```

```python
# Global/models/pix2pixHD_model.py
class Pix2PixHDModel(BaseModel):
    def inference(self, label, inst, image=None, feat=None):
        ...
```
---------------------------------Core components----------------------------------

-----------------------------(d) Other module paths-------------------------------
```text
Global/models/mapping_model.py
Global/data/online_dataset_for_old_photos.py
Face_Enhancement/models/pix2pix_model.py
Global/options/base_options.py
Face_Enhancement/util/util.py
Global/detection_util/util.py
Face_Enhancement/options/base_options.py
Global/detection_models/networks.py
Global/models/NonLocal_feature_mapping_model.py
Face_Enhancement/models/networks/architecture.py
Global/util/visualizer.py
Global/train_mapping.py
Global/train_domain_B.py
Global/train_domain_A.py
```
--------------------------------Other module paths--------------------------------


\end{lstlisting}

\subsection*{E. Detailed Experiment Execution Logs}
\label{appendix:exp-logs}

This appendix presents the detailed execution logs for both the MLE-R and GitTaskBench experiments. These logs document the end-to-end experimental workflows, including task execution traces, error handling, repeated trials for failed steps, and verification of final outputs. The format follows top-tier conference standards, with line widths controlled for readability. All logs are presented in English.

\subsubsection*{E.1 Example Demonstration of 
 MLE-R: Cancer Detection Experiment Log}
The following log outlines the execution of a representative task from the MLE-R suite, specifically the histopathologic cancer detection task:

\begin{lstlisting}[language={},caption={Cancer detection model training log.},label={lst:cancer-detection}]
User: 
I need you to analyze the following code repository and use your programming skills to complete the user task:

Task Description:
Kaggle Task: histopathologic-cancer-detection
Overview:
In this competition, you must create an algorithm to identify metastatic cancer in small image patches taken from larger digital pathology scans...
[Full task and data description omitted for brevity]

Working Directory: 
......

Repository Path: 
......

---

Assistant:
Thank you for the detailed information. I have understood the requirements and the structure of the repository. I will create a PyTorch-based script named `train_and_predict.py` for training and inference.

---

System:
File not found: ......

---

Assistant:
The file does not exist. Let's create it.

---

System:
Error: Function edit_file not found.

---

Assistant:
Sorry, I cannot directly edit files. Let's use Python's built-in functions to create and write the file.

---

User:
[Python code to create `train_and_predict.py` is executed]

---

System:
Code output: File created and content written to ......
File successfully created.

---

Assistant:
Great! The file has been created. Let's run the script to start training and prediction.

---

System:
exitcode: 1 (execution failed)
Code output: Traceback (most recent call last):
File ".../train_and_predict.py", line 195, in <module>
main()
File ".../train_and_predict.py", line 131, in main
train_dataset = CancerDataset(csv_file=train_data, root_dir=TRAIN_DIR, transform=transform)
...
TypeError: argument of type 'method' is not iterable
Using device: cuda

---

Assistant:
I see the issue: we passed a DataFrame instead of a CSV file path to the CancerDataset class. Let's update the code accordingly.

---

[Assistant guides user through code updates using Python scripts to patch the file.]

---

System:
Code output: File .../train_and_predict.py has been updated successfully.

---

Assistant:
Now let's try running the script again.

---

System:
exitcode: 1 (execution failed)
Code output: IndentationError: unexpected indent

---

Assistant:
We introduced an indentation error. Let's fix it with another script.

---

System:
Code output: File .../train_and_predict.py has been updated and indentation fixed.
Using device: cuda
Epoch 1/8, Train Loss: 0.3771, Val Loss: 0.2913, Val AUC: 0.9455
Epoch 2/8, Train Loss: 0.2883, Val Loss: 0.2473, Val AUC: 0.9614
Epoch 3/8, Train Loss: 0.2501, Val Loss: 0.2200, Val AUC: 0.9697
Epoch 4/8, Train Loss: 0.2246, Val Loss: 0.2143, Val AUC: 0.9718
Epoch 5/8, Train Loss: 0.2101, Val Loss: 0.2090, Val AUC: 0.9718
Epoch 6/8, Train Loss: 0.1978, Val Loss: 0.1912, Val AUC: 0.9771
Epoch 7/8, Train Loss: 0.1863, Val Loss: 0.2019, Val AUC: 0.9739
Epoch 8/8, Train Loss: 0.1774, Val Loss: 0.2126, Val AUC: 0.9766
Training and prediction completed. Results saved in 'result_submission.csv'
Result file exists and has the correct format.

---

Assistant:
Excellent! The script ran successfully, trained the model for 8 epochs, and produced a valid result file. Validation AUC reached 0.9766. The workflow included:
- Initial file creation and coding.
- Error handling and iterative code patching.
- Final successful execution and result verification.
\end{lstlisting}

\subsubsection*{E.2 Example Demonstration of 
 GitTaskBench: Old Photo Scratch Removal Experiment Log}
Below is the detailed log for an automated scratch removal task using the DeScratch codebase from the GitTaskBench suite. This log demonstrates the full process, including environment setup, error handling, dependency management, repeated attempts, and final verification:

\begin{lstlisting}[language={},caption={Old photo scratch removal experiment log.},label={lst:descratch}]
User:
Task: Remove scratches from an old photo using the DeScratch repository.
Input: task28/DeScratch_01/.../DeScratch/input_dataset/DeScratch_01_input.jpeg
Output directory: coding_run/task28/DeScratch_01/.../output_result

---

Assistant:
1. Check input image existence.
2. Set up environment variables.
3. Run main script to process image.
4. Verify output.

---

System:
Input file exists but is an image (cannot display directly).

---

Assistant:
Set PYTHONPATH and prepare environment.

---

System:
Environment setup completed.

---

Assistant:
Run main DeScratch processing script via subprocess.

---

System:
exitcode: 1 (execution failed)
Error: FileNotFoundError: No such file or directory: './Global'
Command output: Running Stage 1: Overall restoration

---

Assistant:
Likely not running in the correct directory. Change working directory to repo root and rerun.

---

System:
exitcode: 0 (execution succeeded)
Command output: 
You are using NL + Res
./checkpoints/restoration/VAE_A_quality/latest_net_G.pth not exists yet!
./checkpoints/restoration/VAE_B_scratch/latest_net_G.pth not exists yet!
./checkpoints/restoration/mapping_scratch/latest_net_mapping_net.pth not exists yet!
Running Stage 1: Overall restoration
Finish Stage 1 ...
...
All the processing is done. Please check the results.
Image processing completed successfully.
Output files: []

---

Assistant:
Script ran, but output directory is empty. Warnings indicate missing pretrained model checkpoints.

---

Assistant:
Check output folder structure, input files, and checkpoint existence.

---

System:
Output folder structure:
output_result/
    final_output/
    stage_2_detection_output/
    stage_3_face_output/
    stage_1_restore_output/
        origin/
        input_image/
        restored_image/
No image files found in output folder.
Input folder: ['.ipynb_checkpoints', 'DeScratch_01_input.jpeg']
Checkpoints missing.

---

Assistant:
README indicates pretrained models must be downloaded. Download and extract required checkpoints.

---

System:
Checkpoints downloaded and extracted successfully. Zip files removed.

---

Assistant:
Rerun processing script.

---

System:
exitcode: 0 (execution succeeded)
Command output:
You are using NL + Res
Running Stage 1: Overall restoration
Finish Stage 1 ...
...
All the processing is done. Please check the results.
Image processing completed successfully.
Output files: []

---

Assistant:
Still no output files. Check output folder structure and run.log for clues.

---

System:
Output folder structure:
output_result/
    final_output/
    stage_2_detection_output/
    stage_3_face_output/
    stage_1_restore_output/
        origin/
        input_image/
        restored_image/
Input folder: ['.ipynb_checkpoints', 'DeScratch_01_input.jpeg']
run.log file not found

---

Assistant:
Try to directly call run.py processing steps for more debug output.

---

System:
exitcode: 0 (execution succeeded)
Traceback (most recent call last):
  File ".../DeScratch/Global/detection.py", line 17, in <module>
    from detection_models import networks
  ...
ModuleNotFoundError: No module named 'detection_models.sync_batchnorm'
...
ModuleNotFoundError: No module named 'dlib'
...
Output files: []
stage_1_restore_output: ['origin', 'input_image', 'restored_image']
stage_2_detection_output: []
stage_3_face_output: []

---

Assistant:
Missing Python modules: detection_models.sync_batchnorm, dlib, models.networks.sync_batchnorm.
Install dependencies.

---

System:
All dependencies installed successfully.

---

Assistant:
Prepare input/output folders. Confirm input image exists.

---

System:
Files in input folder: ['.ipynb_checkpoints', 'DeScratch_01_input.jpeg']
Input and output folders prepared.

---

Assistant:
Rerun processing, focus on Stage 1.

---

System:
exitcode: 0 (execution succeeded)
Error executing detection.py: ModuleNotFoundError: No module named 'detection_models.sync_batchnorm'
Test output: You are using NL + Res
Now you are processing DeScratch_01_input.jpeg
Finish Stage 1 ...
output_result/
    final_output/
        DeScratch_01_input.jpeg
    stage_1_restore_output/
        origin/
            DeScratch_01_input.jpeg
        input_image/
            DeScratch_01_input.jpeg
        restored_image/
            DeScratch_01_input.jpeg
        masks/
            mask/
                DeScratch_01_input.png
            input/
                DeScratch_01_input.jpeg

---

Assistant:
Still missing sync_batchnorm. Clone Synchronized-BatchNorm-PyTorch and copy sync_batchnorm.

---

System:
Synchronized BatchNorm module installed successfully.

---

Assistant:
Rerun process, now error: ModuleNotFoundError: No module named 'easydict'.
Install easydict and other dependencies.

---

System:
Additional dependencies installed successfully.

---

Assistant:
Rerun processing script with extra debug output.

---

System:
exitcode: 0 (execution succeeded)
Running Stage 1: Overall restoration
Copied input image to .../masks/input
Created blank mask at .../masks/mask/DeScratch_01_input.png
Executing detection.py --test_path .../input_dataset ...
Command output: initializing the dataloader
model weights loaded
processing DeScratch_01_input.jpeg
Executing test.py --Scratch_and_Quality_restore ...
Command output: You are using NL + Res
Now you are processing DeScratch_01_input..png
Now you are processing DeScratch_01_input.jpeg
Copied DeScratch_01_input..png to final output directory
Copied DeScratch_01_input.jpeg to final output directory
Finish Stage 1 ...
output_result/
    final_output/
        DeScratch_01_input..png
        DeScratch_01_input.jpeg
    stage_1_restore_output/
        origin/
            DeScratch_01_input..png
            DeScratch_01_input.jpeg
        input_image/
            DeScratch_01_input..png
            DeScratch_01_input.jpeg
        restored_image/
            DeScratch_01_input..png
            DeScratch_01_input.jpeg
        masks/
            mask/
                DeScratch_01_input.png
                DeScratch_01_input..png
            input/
                DeScratch_01_input..png
                DeScratch_01_input.jpeg

---

Assistant:
Processing succeeded. Compare input and output images.

---

System:
Comparison image saved as 'image_comparison.png'
Input image size: (559, 878)
Output image size: (556, 876)
Input image format: JPEG
Output image format: JPEG

---

Assistant:
Processing completed. Input and output images are available for visual comparison. The pipeline handled errors through iterative debugging, resolved all missing dependencies, and produced restored images as required.
\end{lstlisting}

\noindent These logs collectively demonstrate the robustness and reproducibility of our experimental framework. All steps, including failures and subsequent recovery, are documented to ensure transparency and facilitate future benchmarking.

\subsection*{F. Details of Module‑Level Scoring Features}
\label{app:feature-details}

\textbf{Dependency.} As shown in Algorithm \ref{alg:pagerank}, to measure a module's centrality in the module dependency graph, we apply Personalized PageRank, emphasizing the target module with a personalized vector. The algorithm iteratively computes the PageRank score using a damping factor of  \(\alpha=0.85\), and the resulting score is scaled for comparability with other metrics.

\textbf{Complexity.} The feature calculates the code complexity of a module by evaluating two core factors: the frequency of branching and looping structures, and the depth of function nesting. The occurrence frequency of branching and looping structures (such as \texttt{if}, \texttt{for}, \texttt{while}, and \texttt{except}) is normalized to prevent the complexity score from becoming excessively high. The function nesting depth is quantified based on the indentation level, and the resulting score is scaled by a proportional factor. The final complexity score is a composite of these two factors, with each factor's score normalized to a maximum value of 1.0, providing a comprehensive and balanced measure of the module's code complexity, as described in Algorithm~\ref{alg:complexity}.

\textbf{Usage.} This feature calculates the number of times other modules import the current module, as well as the number of times the functions and classes within the module are called, using these metrics as indicators of the module's significance.

\textbf{Semantic.} we conduct a semantic importance analysis based on the naming conventions of modules to identify key files. Commonly recognized names such as \texttt{\_\_init\_\_}, \texttt{app}, \texttt{settings}, \texttt{config}, \texttt{utils}, and \texttt{constants} are considered to be of higher significance. In addition to this, we examine the presence of crucial semantic keywords within the code, including terms like \texttt{main}, \texttt{core}, \texttt{engine}, \texttt{api}, \texttt{service}, \texttt{processor}, \texttt{factory}, \texttt{builder}, \texttt{scheduler}, and \texttt{config}. The occurrence of these keywords in module or function names serves as an important factor in determining the module's relevance and is used as a scoring criterion in our analysis.

\textbf{Doc.} This feature calculates a score based on the existence and quality of the module's docstring to assess the completeness of the module's documentation. The score is initially determined by the length of the docstring. Additionally, the feature checks for the presence of specific sections within the docstring, such as descriptions of parameters, return values, and examples. If the first two sections are present, the score increases for each, and if examples are included, the score also increases. This approach provides a comprehensive evaluation of the documentation's completeness.

\textbf{Git.} This feature calculates whether the file is part of a Git repository or not. If the file is not in a Git repository, a score of 0.0 is returned. Next, it counts the number of commits associated with the file. The score is initially based on the number of commits. The feature also takes into account the time of the last commit, and recently modified files are considered more important, with the time factor accounting for 30\% of the final score. The final importance score is a weighted combination of the number of submissions and the time-of-submission score to ensure that both the frequency of modifications and the freshness of the file are taken into account.

\begin{table}[htbp]
    \centering
    \footnotesize
    \renewcommand{\arraystretch}{1.1}
    \setlength{\tabcolsep}{3pt}
    \caption{Token Usage by Agent Across LLMs on MLE-R (in Millions of Tokens).}
    \begin{tabular}{l c c c}
        \toprule
        \textbf{Agent} & \textbf{GPT‑4o (M)} & \textbf{Claude 3.5 (M)} & \textbf{Deepseek v3 (M)} \\
        \midrule
        Openhands    & 5.26 & 4.73 & 4.93 \\
        SWE‑Agent    & 1.29 & 1.51 & 1.48 \\
        RepoMaster   & 0.36 & 0.32 & 0.18 \\
        \bottomrule
    \end{tabular}
    \label{tab:agent_token_usage_mle}
\end{table}

\begin{algorithm}[htbp]
\caption{Calculate Personalized PageRank Score for a Module}
\label{alg:pagerank}
\begin{algorithmic}[1]
\STATE \textbf{Input:} Module dependency graph $G_{m}$, target module $m$
\STATE \textbf{Output:} Personalized PageRank score $score$
\STATE $score \leftarrow 0.0$
\IF{$|V(G_{m})| > 0$}
    \STATE $personalization[n] \leftarrow 1.0$ for all $n \in V(G_{m})$
    \STATE $personalization[m] \leftarrow 2.0$
    \STATE $pagerank \leftarrow \text{PageRank}(G_{m}, \alpha=0.85, personalization)$
    \STATE $score \leftarrow pagerank[moduleId] \times 10$
\ENDIF
\RETURN $score$
\end{algorithmic}
\end{algorithm}

\begin{algorithm}[htbp]
\caption{Analyze Code Complexity of a Module}
\label{alg:complexity}
\begin{algorithmic}[1]
\STATE \textbf{Input:} target module $m$
\STATE \textbf{Output:} Code complexity score $score$
\STATE $score \gets 0.0$
\STATE $content \gets m.content$

\STATE $branchCount \gets \sum_{k \in \{\emph{if}, \emph{for}, \emph{while}, \emph{except}\}} \text{count}(k, \text{in } content)$

\STATE $score \gets \min(branchCount / 50.0, 1.0)$

\STATE $matches \gets \text{count}(\emph{def}, \text{in } content)$
\IF{$matches \neq \emptyset$}
    \STATE $indentLevel \gets \max(matches) / 4$
    \STATE $score \gets score + \min(indentLevel / 5.0, 1.0) \times 0.3$
\ENDIF
\RETURN $score$
\end{algorithmic}
\end{algorithm}

\subsection*{G. Detailed Experimental Results and Analysis}
\subsubsection*{G.1 Detailed Results on MLE-R}
To ensure transparency and facilitate reproducibility, we present the complete raw results of our experiments in this section. Tables~\ref{tab:RepoMaster_gpt4o_results},~\ref{tab:RepoMaster_claude_results} and~\ref{tab:RepoMaster_deepseek_results} present RepoMaster's scores on all 22 MLE‑R tasks when paired with GPT‑4o, Claude 3.5, and Deepseek v3, respectively, reporting for each task the achieved score, relevant medal‑category thresholds, and binary indicators of whether performance criteria are met, thereby enabling detailed comparisons between models and supporting further analysis of their capabilities. Additionally, Tables~\ref{tab:openhands_deepseek_results_updated} and~\ref{tab:sweagent_claude_results} show the results of Openhands and SWE‑Agent on MLE‑R with Claude 3.5, respectively.

Table~\ref{tab:agent_token_usage_mle} presents the token usage of each agent across GPT‑4o, Claude 3.5, and Deepseek v3 when handling tasks from the MLE‑R dataset. It shows that Openhands consumes the most tokens (around 5M per task), SWE‑Agent uses a moderate amount (1.3–1.5M), and RepoMaster is by far the most efficient (0.18–0.36M), with all three agents exhibiting a slight reduction in token usage when moving from GPT‑4o to Claude-3.5 to Deepseek V3; this pattern highlights how model‑specific tokenization and interaction protocols affect consumption, with RepoMaster’s lightweight design offering significant cost savings for large‑scale MLE‑R tasks.

\subsubsection*{G.2 Detailed Results on GitTaskBench}
We report the detailed results(especially the number of tokens) of different frameworks with different models on GitTaskBench in Tables~\ref{tab:gitbench_detail_comparison}.

\subsubsection*{G.3 Ablation Study}
\begin{figure}[htbp]
    \centering
    \includegraphics[width=1\textwidth]{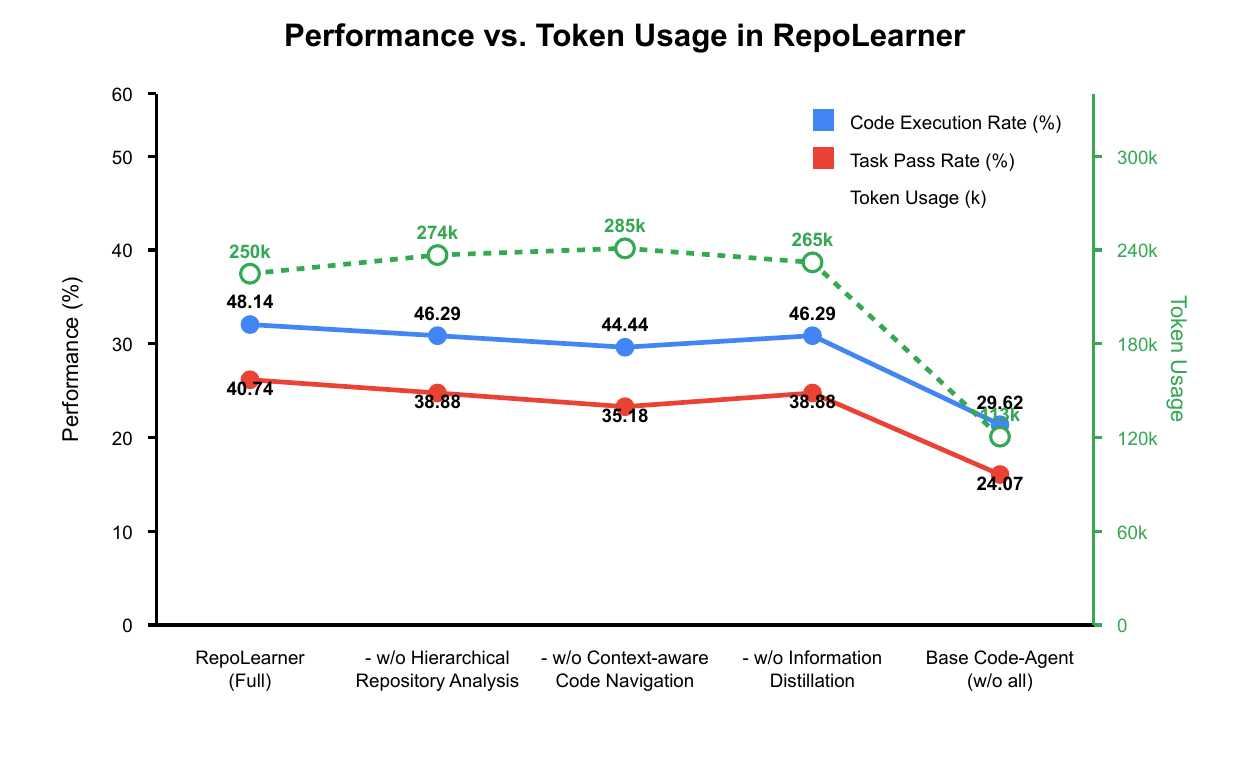}
    \caption{Performance comparison of different ablation configurations in RepoMaster.}
    \label{fig:ablation_bar}
\end{figure}

To more intuitively represent the role of each core component, we present the results of the Ablation Study visually, as shown in Figure~\ref{fig:ablation_bar}.

\begin{table}[h]
    \centering
    \renewcommand{\arraystretch}{1.2}
    \setlength{\tabcolsep}{8pt}
    \caption{Comparison of Different Frameworks on GitTaskBench.}
    \resizebox{\textwidth}{!}{
    \begin{tabular}{l l c c c}
        \toprule
        \textbf{Framework} & \textbf{LLM} & \textbf{Execution Completion Rate (\%)} & \textbf{Task Pass Rate  (\%)} & \textbf{\#Total Tokens} \\
        \midrule

        \multirow{3}{*}{SWE-Agent}
            & \multicolumn{1}{>{\columncolor{gptbg}}l}{GPT-4o} & \multicolumn{1}{>{\columncolor{gptbg}}c}{29.63} & \multicolumn{1}{>{\columncolor{gptbg}}c}{9.26} & \multicolumn{1}{>{\columncolor{gptbg}}c}{308155} \\
            & \multicolumn{1}{>{\columncolor{claudebg}}l}{Claude 3.5} & \multicolumn{1}{>{\columncolor{claudebg}}c}{44.44} & \multicolumn{1}{>{\columncolor{claudebg}}c}{14.81} & \multicolumn{1}{>{\columncolor{claudebg}}c}{330676} \\
            & \multicolumn{1}{>{\columncolor{llamabg}}l}{DeepSeek V3} & \multicolumn{1}{>{\columncolor{llamabg}}c}{29.63} & \multicolumn{1}{>{\columncolor{llamabg}}c}{9.26} & \multicolumn{1}{>{\columncolor{llamabg}}c}{265538} \\
        \midrule
        
        \multirow{3}{*}{OpenHands}
            & \multicolumn{1}{>{\columncolor{gptbg}}l}{GPT-4o} & \multicolumn{1}{>{\columncolor{gptbg}}c}{37.04} & \multicolumn{1}{>{\columncolor{gptbg}}c}{14.81} & \multicolumn{1}{>{\columncolor{gptbg}}c}{1195312} \\
            & \multicolumn{1}{>{\columncolor{claudebg}}l}{Claude 3.5} & \multicolumn{1}{>{\columncolor{claudebg}}c}{48.15} & \multicolumn{1}{>{\columncolor{claudebg}}c}{24.07} & \multicolumn{1}{>{\columncolor{claudebg}}c}{3094878} \\
            & \multicolumn{1}{>{\columncolor{llamabg}}l}{DeepSeek V3} & \multicolumn{1}{>{\columncolor{llamabg}}c}{42.59} & \multicolumn{1}{>{\columncolor{llamabg}}c}{16.67} & \multicolumn{1}{>{\columncolor{llamabg}}c}{7662130} \\
        \midrule
        
        \multirow{3}{*}{RepoMaster}
            & \multicolumn{1}{>{\columncolor{gptbg}}l}{GPT-4o} & \multicolumn{1}{>{\columncolor{gptbg}}c}{48.14} & \multicolumn{1}{>{\columncolor{gptbg}}c}{40.74} & \multicolumn{1}{>{\columncolor{gptbg}}c}{250245} \\
            & \multicolumn{1}{>{\columncolor{claudebg}}l}{Claude 3.5} & \multicolumn{1}{>{\columncolor{claudebg}}c}{75.92} & \multicolumn{1}{>{\columncolor{claudebg}}c}{62.96} & \multicolumn{1}{>{\columncolor{claudebg}}c}{154913} \\
            & \multicolumn{1}{>{\columncolor{llamabg}}l}{DeepSeek V3} & \multicolumn{1}{>{\columncolor{llamabg}}c}{61.11} & \multicolumn{1}{>{\columncolor{llamabg}}c}{44.44} & \multicolumn{1}{>{\columncolor{llamabg}}c}{255998} \\
        \bottomrule
    \end{tabular}
    }
    \label{tab:gitbench_detail_comparison}
\end{table}

\begin{table}[htbp]
    \centering
    \renewcommand{\arraystretch}{1.2}
    \setlength{\tabcolsep}{4pt}
    \caption{RepoMaster with GPT-4o: Raw Results.}
    \resizebox{\textwidth}{!}{
    \begin{tabular}{l c c c c c c c c c c c}
        \toprule
        \textbf{Competition} & \textbf{Score} & \textbf{Gold} & \textbf{Silver} & \textbf{Bronze} & \textbf{Median} & \textbf{Any} & \textbf{Gold} & \textbf{Silver} & \textbf{Bronze} & \textbf{Above} & \textbf{Valid} \\
        & & \textbf{Threshold} & \textbf{Threshold} & \textbf{Threshold} & \textbf{Threshold} & \textbf{Medal} & \textbf{Medal} & \textbf{Medal} & \textbf{Medal} & \textbf{Median} & \textbf{Sub.} \\
        \midrule
        aerial-cactus-identification & 1.00000 & 1.00000 & 1.00000 & 1.00000 & 0.99910 & TRUE & TRUE & FALSE & FALSE & TRUE & TRUE \\
        aptos2019-blindness-detection & 0.89501 & 0.93051 & 0.91965 & 0.91449 & 0.88891 & FALSE & FALSE & FALSE & FALSE & TRUE & TRUE \\
        chaii-hindi-and-tamil-question-answering & 0.68616 & 0.76305 & 0.73984 & 0.73725 & 0.72756 & FALSE & FALSE & FALSE & FALSE & FALSE & TRUE \\
        denoising-dirty-documents & - & 0.01794 & 0.02609 & 0.04517 & 0.07325 & FALSE & FALSE & FALSE & FALSE & FALSE & FALSE \\
        dog-breed-identification & 0.72540 & 0.00050 & 0.00539 & 0.04598 & 0.47205 & FALSE & FALSE & FALSE & FALSE & FALSE & TRUE \\
        dogs-vs-cats-redux-kernels-edition & 0.10353 & 0.03882 & 0.05038 & 0.06127 & 0.12216 & FALSE & FALSE & FALSE & FALSE & TRUE & TRUE \\
        jigsaw-toxic-comment-classification-challenge & 0.98639 & 0.98740 & 0.98668 & 0.98639 & 0.98079 & TRUE & FALSE & FALSE & TRUE & TRUE & TRUE \\
        histopathologic-cancer-detection & 0.99112 & 0.98350 & 0.97980 & 0.97380 & 0.94770 & TRUE & TRUE & FALSE & FALSE & TRUE & TRUE \\
        leaf-classification & 0.04422 & 0.00000 & 0.00791 & 0.01526 & 0.10835 & FALSE & FALSE & FALSE & FALSE & TRUE & TRUE \\
        learning-agency-lab-automated-essay-scoring-2 & 0.78242 & 0.83583 & 0.83518 & 0.83471 & 0.82827 & FALSE & FALSE & FALSE & FALSE & FALSE & TRUE \\
        lmsys-chatbot-arena & 1.08268 & 0.98392 & 0.99277 & 1.00283 & 1.05212 & FALSE & FALSE & FALSE & FALSE & FALSE & TRUE \\
        mlsp-2013-birds & - & 0.93527 & 0.90038 & 0.87372 & 0.86572 & FALSE & FALSE & FALSE & FALSE & FALSE & FALSE \\
        new-york-city-taxi-fare-prediction & 4.76987 & 2.83377 & 2.88191 & 2.92371 & 3.59741 & FALSE & FALSE & FALSE & FALSE & FALSE & TRUE \\
        nomad2018-predict-transparent-conductors & 0.06692 & 0.05589 & 0.06229 & 0.06582 & 0.06988 & FALSE & FALSE & FALSE & FALSE & TRUE & TRUE \\
        plant-pathology-2020-fgvc7 & 0.94049 & 0.97836 & 0.97465 & 0.97361 & 0.94852 & FALSE & FALSE & FALSE & FALSE & FALSE & TRUE \\
        random-acts-of-pizza & 0.63818 & 0.97908 & 0.76482 & 0.69210 & 0.59960 & FALSE & FALSE & FALSE & FALSE & TRUE & TRUE \\
        spooky-author-identification & - &0.16506 & 0.26996 & 0.29381 & 0.41879 & FALSE & FALSE & FALSE & FALSE & FALSE & FALSE \\
        tabular-playground-series-dec-2021 & 0.95711 & 0.95660 & 0.95658 & 0.95658 & 0.95342 & TRUE & TRUE & FALSE & FALSE & TRUE & TRUE \\
        ranzcr-clip-catheter-line-classification & 0.84915 & 0.97357 & 0.97152 & 0.97090 & 0.96750 & FALSE & FALSE & FALSE & FALSE & FALSE & TRUE \\
        tabular-playground-series-may-2022 & 0.95056 & 0.99823 & 0.99822 & 0.99818 & 0.97268 & FALSE & FALSE & FALSE & FALSE & FALSE & TRUE \\
        text-normalization-challenge-russian-language & 0.00000 & 0.99012 & 0.98232 & 0.97592 & 0.97591 & FALSE & FALSE & FALSE & FALSE & FALSE & TRUE \\
        tgs-salt-identification-challenge & 0.52210 & 0.89097 & 0.87163 & 0.85918 & 0.81665 & FALSE & FALSE & FALSE & FALSE & FALSE & TRUE \\
        \bottomrule
    \end{tabular}
    }
    \label{tab:RepoMaster_gpt4o_results}
\end{table}

\begin{table}[htbp]
    \centering
    \renewcommand{\arraystretch}{1.2}
    \setlength{\tabcolsep}{4pt}
    \caption{RepoMaster with Claude 3.5: Raw Results.}
    \resizebox{\textwidth}{!}{
    \begin{tabular}{l c c c c c c c c c c c}
        \toprule
        \textbf{Competition} & \textbf{Score} & \textbf{Gold} & \textbf{Silver} & \textbf{Bronze} & \textbf{Median} & \textbf{Any} & \textbf{Gold} & \textbf{Silver} & \textbf{Bronze} & \textbf{Above} & \textbf{Valid} \\
        & & \textbf{Threshold} & \textbf{Threshold} & \textbf{Threshold} & \textbf{Threshold} & \textbf{Medal} & \textbf{Medal} & \textbf{Medal} & \textbf{Medal} & \textbf{Median} & \textbf{Sub.} \\
        \midrule
        aerial-cactus-identification & 1.00000 & 1.00000 & 1.00000 & 1.00000 & 0.99910 & TRUE & TRUE & FALSE & FALSE & TRUE & TRUE \\
        aptos2019-blindness-detection & 0.90685 & 0.93051 & 0.91965 & 0.91449 & 0.88891 & FALSE & FALSE & FALSE & FALSE & TRUE & TRUE \\
        chaii-hindi-and-tamil-question-answering & 0.67455 & 0.76305 & 0.73984 & 0.73725 & 0.72756 & FALSE & FALSE & FALSE & FALSE & FALSE & TRUE \\
        denoising-dirty-documents & - & 0.01794 & 0.02609 & 0.04517 & 0.07325 & FALSE & FALSE & FALSE & FALSE & FALSE & FALSE \\
        dog-breed-identification & 0.58826 & 0.00050 & 0.00539 & 0.04598 & 0.47205 & FALSE & FALSE & FALSE & FALSE & FALSE & TRUE \\
        dogs-vs-cats-redux-kernels-edition & 0.06683 & 0.03882 & 0.05038 & 0.06127 & 0.12216 & FALSE & FALSE & FALSE & FALSE & TRUE & TRUE \\
        jigsaw-toxic-comment-classification-challenge & 0.98929 & 0.98740 & 0.98668 & 0.98639 & 0.98079 & TRUE & TRUE & FALSE & FALSE & TRUE & TRUE \\
        histopathologic-cancer-detection & 0.99254 & 0.98350 & 0.97980 & 0.97380 & 0.94770 & TRUE & TRUE & FALSE & FALSE & TRUE & TRUE \\
        leaf-classification & 2.02279 & 0.00000 & 0.00791 & 0.01526 & 0.10835 & FALSE & FALSE & FALSE & FALSE & FALSE & TRUE \\
        learning-agency-lab-automated-essay-scoring-2 & 0.78964 & 0.83583 & 0.83518 & 0.83471 & 0.82827 & FALSE & FALSE & FALSE & FALSE & FALSE & TRUE \\
        lmsys-chatbot-arena & 2.26734 & 0.98392 & 0.99277 & 1.00283 & 1.05212 & FALSE & FALSE & FALSE & FALSE & FALSE & TRUE \\
        mlsp-2013-birds & 0.65981 & 0.93527 & 0.90038 & 0.87372 & 0.86572 & FALSE & FALSE & FALSE & FALSE & FALSE & TRUE \\
        new-york-city-taxi-fare-prediction & 5.12441 & 2.83377 & 2.88191 & 2.92371 & 3.59741 & FALSE & FALSE & FALSE & FALSE & FALSE & TRUE \\
        nomad2018-predict-transparent-conductors & 0.06479 & 0.05589 & 0.06229 & 0.06582 & 0.06988 & TRUE & FALSE & FALSE & TRUE & TRUE & TRUE \\
        plant-pathology-2020-fgvc7 & 0.98888 & 0.97836 & 0.97465 & 0.97361 & 0.94852 & TRUE & TRUE & FALSE & FALSE & TRUE & TRUE \\
        random-acts-of-pizza & 0.63403 & 0.97908 & 0.76482 & 0.69210 & 0.59960 & FALSE & FALSE & FALSE & FALSE & TRUE & TRUE \\
        spooky-author-identification & 0.50796 & 0.16506 & 0.26996 & 0.29381 & 0.41879 & FALSE & FALSE & FALSE & FALSE & FALSE & TRUE \\
        tabular-playground-series-dec-2021 & 0.96227 & 0.95660 & 0.95658 & 0.95658 & 0.95342 & TRUE & TRUE & FALSE & FALSE & TRUE & TRUE \\
        ranzcr-clip-catheter-line-classification & 0.91278 & 0.97357 & 0.97152 & 0.97090 & 0.96750 & FALSE & FALSE & FALSE & FALSE & FALSE & TRUE \\
        tabular-playground-series-may-2022 & 0.99373 & 0.99823 & 0.99822 & 0.99818 & 0.97268 & FALSE & FALSE & FALSE & FALSE & TRUE & TRUE \\
        text-normalization-challenge-russian-language & 0.18164 & 0.99012 & 0.98232 & 0.97592 & 0.97591 & FALSE & FALSE & FALSE & FALSE & FALSE & TRUE \\
        tgs-salt-identification-challenge & 0.52210 & 0.89097 & 0.87163 & 0.85918 & 0.81665 & FALSE & FALSE & FALSE & FALSE & FALSE & TRUE \\
        \bottomrule
    \end{tabular}
    }
    \label{tab:RepoMaster_claude_results}
\end{table}

\begin{table}[htbp]
    \centering
    \renewcommand{\arraystretch}{1.2}
    \setlength{\tabcolsep}{4pt}
    \caption{RepoMaster with Deepseek V3: Raw Results.}
    \resizebox{\textwidth}{!}{
    \begin{tabular}{l c c c c c c c c c c c}
        \toprule
        \textbf{Competition} & \textbf{Score} & \textbf{Gold} & \textbf{Silver} & \textbf{Bronze} & \textbf{Median} & \textbf{Any} & \textbf{Gold} & \textbf{Silver} & \textbf{Bronze} & \textbf{Above} & \textbf{Valid} \\
        & & \textbf{Threshold} & \textbf{Threshold} & \textbf{Threshold} & \textbf{Threshold} & \textbf{Medal} & \textbf{Medal} & \textbf{Medal} & \textbf{Medal} & \textbf{Median} & \textbf{Sub.} \\
        \midrule
        aerial-cactus-identification & 0.99988 & 1.00000 & 1.00000 & 1.00000 & 0.99910 & FALSE & FALSE & FALSE & FALSE & TRUE  & TRUE  \\
        aptos2019-blindness-detection & 0.87334 & 0.93051 & 0.91965 & 0.91449 & 0.88891 & FALSE & FALSE & FALSE & FALSE & FALSE & TRUE  \\
        chaii-hindi-and-tamil-question-answering & 0.31169 & 0.76305 & 0.73984 & 0.73725 & 0.72756 & FALSE & FALSE & FALSE & FALSE & FALSE & TRUE  \\
        denoising-dirty-documents & -       & 0.01794 & 0.02609 & 0.04517 & 0.07325 & FALSE & FALSE & FALSE & FALSE & FALSE & FALSE \\
        dog-breed-identification & 0.84673 & 0.00050 & 0.00539 & 0.04598 & 0.47205 & FALSE & FALSE & FALSE & FALSE & FALSE & TRUE  \\
        dogs-vs-cats-redux-kernels-edition & 0.04717 & 0.03882 & 0.05038 & 0.06127 & 0.12216 & TRUE  & FALSE & TRUE  & FALSE & TRUE  & TRUE  \\
        histopathologic-cancer-detection & 0.99300 & 0.98350 & 0.97980 & 0.97380 & 0.94770 & TRUE  & TRUE  & FALSE & FALSE & TRUE  & TRUE  \\
        jigsaw-toxic-comment-classification-challenge & 0.96374 & 0.98740 & 0.98668 & 0.98639 & 0.98079 & FALSE & FALSE & FALSE & FALSE & FALSE & TRUE  \\
        leaf-classification & 0.07526 & 0.00000 & 0.00791 & 0.01526 & 0.10835 & FALSE & FALSE & FALSE & FALSE & TRUE  & TRUE  \\
        learning-agency-lab-automated-essay-scoring-2 & 0.89558 & 0.83583 & 0.83518 & 0.83471 & 0.82827 & TRUE  & TRUE  & FALSE & FALSE & TRUE  & TRUE  \\
        lmsys-chatbot-arena & 1.09199 & 0.98392 & 0.99277 & 1.00283 & 1.05212 & FALSE & FALSE & FALSE & FALSE & FALSE & TRUE  \\
        mlsp-2013-birds & -       & 0.93527 & 0.90038 & 0.87372 & 0.86572 & FALSE & FALSE & FALSE & FALSE & FALSE & FALSE \\
        new-york-city-taxi-fare-prediction & 1064.68701 & 2.83377 & 2.88191 & 2.92371 & 3.59741 & FALSE & FALSE & FALSE & FALSE & FALSE & TRUE  \\
        nomad2018-predict-transparent-conductors & 0.06468 & 0.05589 & 0.06229 & 0.06582 & 0.06988 & TRUE  & FALSE & FALSE & TRUE  & TRUE  & TRUE  \\
        plant-pathology-2020-fgvc7 & 0.92653 & 0.97836 & 0.97465 & 0.97361 & 0.94852 & FALSE & FALSE & FALSE & FALSE & FALSE & TRUE  \\
        random-acts-of-pizza & 0.59387 & 0.97908 & 0.76482 & 0.69210 & 0.59960 & FALSE & FALSE & FALSE & FALSE & FALSE & TRUE  \\
        ranzcr-clip-catheter-line-classification & -       & 0.97357 & 0.97152 & 0.97090 & 0.96750 & FALSE & FALSE & FALSE & FALSE & FALSE & FALSE \\
        spooky-author-identification & 0.62857 & 0.16506 & 0.26996 & 0.29381 & 0.41879 & FALSE & FALSE & FALSE & FALSE & FALSE & TRUE  \\
        tabular-playground-series-dec-2021 & 0.96084 & 0.95660 & 0.95658 & 0.95658 & 0.95342 & TRUE  & TRUE  & FALSE & FALSE & TRUE  & TRUE  \\
        tabular-playground-series-may-2022 & 0.99161 & 0.99823 & 0.99822 & 0.99818 & 0.97268 & FALSE & FALSE & FALSE & FALSE & TRUE  & TRUE  \\
        text-normalization-challenge-russian-language & 0.01451 & 0.99012 & 0.98232 & 0.97592 & 0.97591 & FALSE & FALSE & FALSE & FALSE & FALSE & TRUE  \\
        tgs-salt-identification-challenge & 0.50440 & 0.89097 & 0.87163 & 0.85918 & 0.81665 & FALSE & FALSE & FALSE & FALSE & FALSE & TRUE  \\
        \bottomrule
    \end{tabular}
    }
    \label{tab:RepoMaster_deepseek_results}
\end{table}

\begin{table}[htbp]
    \centering
    \renewcommand{\arraystretch}{1.2}
    \setlength{\tabcolsep}{4pt}
    \caption{Openhands with Claude 3.5: Raw  Results.}
    \resizebox{\textwidth}{!}{
    \begin{tabular}{l c c c c c c c c c c c}
        \toprule
        \textbf{Competition} & \textbf{Score} & \textbf{Gold} & \textbf{Silver} & \textbf{Bronze} & \textbf{Median} & \textbf{Any} & \textbf{Gold} & \textbf{Silver} & \textbf{Bronze} & \textbf{Above} & \textbf{Valid} \\
        & & \textbf{Threshold} & \textbf{Threshold} & \textbf{Threshold} & \textbf{Threshold} & \textbf{Medal} & \textbf{Medal} & \textbf{Medal} & \textbf{Medal} & \textbf{Median} & \textbf{Sub.} \\
        \midrule
        aerial-cactus-identification            & 0.99951  & 1.00000 & 1.00000 & 1.00000 & 0.99910 & FALSE & FALSE & FALSE & FALSE & FALSE & TRUE  \\
        aptos2019-blindness-detection           & –        & 0.93051 & 0.91965 & 0.91449 & 0.88891 & FALSE & FALSE & FALSE & FALSE & FALSE & FALSE \\
        chaii-hindi-and-tamil-question-answering & –       & 0.76305 & 0.73984 & 0.73725 & 0.72756 & FALSE & FALSE & FALSE & FALSE & FALSE & FALSE \\
        denoising-dirty-documents               & –        & 0.01794 & 0.02609 & 0.04517 & 0.07325 & FALSE & FALSE & FALSE & FALSE & FALSE & FALSE \\
        dog-breed-identification                & –        & 0.00050 & 0.00539 & 0.04598 & 0.47205 & FALSE & FALSE & FALSE & FALSE & FALSE & FALSE \\
        dogs-vs-cats-redux-kernels-edition      & –        & 0.03882 & 0.05038 & 0.06127 & 0.12216 & FALSE & FALSE & FALSE & FALSE & FALSE & FALSE \\
        histopathologic-cancer-detection        & –        & 0.98350 & 0.97980 & 0.97380 & 0.94770 & FALSE & FALSE & FALSE & FALSE & FALSE & FALSE \\
        jigsaw-toxic-comment-classification-challenge & – & 0.98740 & 0.98668 & 0.98639 & 0.98079 & FALSE & FALSE & FALSE & FALSE & FALSE & FALSE \\
        leaf-classification                     & 4.59036  & 0.00000 & 0.00791 & 0.01526 & 0.10835 & FALSE & FALSE & FALSE & FALSE & FALSE & TRUE  \\
        learning-agency-lab-automated-essay-scoring-2 & 0.00000 & 0.83583 & 0.83518 & 0.83471 & 0.82827 & FALSE & FALSE & FALSE & FALSE & FALSE & TRUE  \\
        lmsys-chatbot-arena                     & –        & 0.98392 & 0.99277 & 1.00283 & 1.05212 & FALSE & FALSE & FALSE & FALSE & FALSE & FALSE \\
        mlsp-2013-birds                         & –        & 0.93527 & 0.90038 & 0.87372 & 0.86572 & FALSE & FALSE & FALSE & FALSE & FALSE & FALSE \\
        new-york-city-taxi-fare-prediction      & 13.42166 & 2.83377 & 2.88191 & 2.92371 & 3.59741 & FALSE & FALSE & FALSE & FALSE & FALSE & TRUE  \\
        nomad2018-predict-transparent-conductors & 0.23121 & 0.05589 & 0.06229 & 0.06582 & 0.06988 & FALSE & FALSE & FALSE & FALSE & FALSE & TRUE  \\
        plant-pathology-2020-fgvc7              & –        & 0.97836 & 0.97465 & 0.97361 & 0.94852 & FALSE & FALSE & FALSE & FALSE & FALSE & FALSE \\
        random-acts-of-pizza                    & 0.54052  & 0.97908 & 0.76482 & 0.69210 & 0.59960 & FALSE & FALSE & FALSE & FALSE & FALSE & TRUE  \\
        ranzcr-clip-catheter-line-classification & –       & 0.97357 & 0.97152 & 0.97090 & 0.96750 & FALSE & FALSE & FALSE & FALSE & FALSE & FALSE \\
        spooky-author-identification            & 0.92421  & 0.16506 & 0.26996 & 0.29381 & 0.41879 & FALSE & FALSE & FALSE & FALSE & FALSE & TRUE  \\
        tabular-playground-series-dec-2021      & 0.95837  & 0.95660 & 0.95658 & 0.95658 & 0.95342 & TRUE  & TRUE  & FALSE & FALSE & TRUE  & TRUE  \\
        tabular-playground-series-may-2022      & 0.93330  & 0.99823 & 0.99822 & 0.99818 & 0.97268 & FALSE & FALSE & FALSE & FALSE & FALSE & TRUE  \\
        text-normalization-challenge-english-language & – & 0.99012 & 0.98232 & 0.97592 & 0.97591 & FALSE & FALSE & FALSE & FALSE & FALSE & FALSE \\
        tgs-salt-identification-challenge       & 0.44300  & 0.89097 & 0.87163 & 0.85918 & 0.81665 & FALSE & FALSE & FALSE & FALSE & FALSE & TRUE  \\
        \bottomrule
    \end{tabular}
    }
    \label{tab:openhands_deepseek_results_updated}
\end{table}

\begin{table}[htbp]
    \centering
    \renewcommand{\arraystretch}{1.2}
    \setlength{\tabcolsep}{4pt}
    \caption{SWE-Agent with Claude 3.5: Raw  Results.}
    \resizebox{\textwidth}{!}{
    \begin{tabular}{l c c c c c c c c c c c}
        \toprule
        \textbf{Competition} & \textbf{Score} & \textbf{Gold} & \textbf{Silver} & \textbf{Bronze} & \textbf{Median} & \textbf{Any} & \textbf{Gold} & \textbf{Silver} & \textbf{Bronze} & \textbf{Above} & \textbf{Valid} \\
        & & \textbf{Threshold} & \textbf{Threshold} & \textbf{Threshold} & \textbf{Threshold} & \textbf{Medal} & \textbf{Medal} & \textbf{Medal} & \textbf{Medal} & \textbf{Median} & \textbf{Sub.} \\
        \midrule
        aerial-cactus-identification            & 0.99917 & 1.00000 & 1.00000 & 1.00000 & 0.99910 & FALSE & FALSE & FALSE & FALSE & TRUE  & TRUE  \\
        aptos2019-blindness-detection           & -       & 0.93051 & 0.91965 & 0.91449 & 0.88891 & FALSE & FALSE & FALSE & FALSE & FALSE & FALSE \\
        chaii-hindi-and-tamil-question-answering & 0.04985 & 0.76305 & 0.73984 & 0.73725 & 0.72756 & FALSE & FALSE & FALSE & FALSE & FALSE & TRUE  \\
        denoising-dirty-documents               & -       & 0.01794 & 0.02609 & 0.04517 & 0.07325 & FALSE & FALSE & FALSE & FALSE & FALSE & FALSE \\
        dog-breed-identification                & -       & 0.00050 & 0.00539 & 0.04598 & 0.47205 & FALSE & FALSE & FALSE & FALSE & FALSE & FALSE \\
        dogs-vs-cats-redux-kernels-edition      & -       & 0.03882 & 0.05038 & 0.06127 & 0.12216 & FALSE & FALSE & FALSE & FALSE & FALSE & FALSE \\
        histopathologic-cancer-detection        & -       & 0.98350 & 0.97980 & 0.97380 & 0.94770 & FALSE & FALSE & FALSE & FALSE & FALSE & FALSE \\
        jigsaw-toxic-comment-classification-challenge & 0.97251 & 0.98740 & 0.98668 & 0.98639 & 0.98079 & FALSE & FALSE & FALSE & FALSE & FALSE & TRUE  \\
        leaf-classification                     & 2.74648 & 0.00000 & 0.00791 & 0.01526 & 0.10835 & FALSE & FALSE & FALSE & FALSE & FALSE & TRUE  \\
        learning-agency-lab-automated-essay-scoring-2 & -   & 0.83583 & 0.83518 & 0.83471 & 0.82827 & FALSE & FALSE & FALSE & FALSE & FALSE & FALSE \\
        lmsys-chatbot-arena                     & -       & 0.98392 & 0.99277 & 1.00283 & 1.05212 & FALSE & FALSE & FALSE & FALSE & FALSE & FALSE \\
        mlsp-2013-birds                         & 0.69828 & 0.93527 & 0.90038 & 0.87372 & 0.86572 & FALSE & FALSE & FALSE & FALSE & FALSE & TRUE  \\
        new-york-city-taxi-fare-prediction      & -       & 2.83377 & 2.88191 & 2.92371 & 3.59741 & FALSE & FALSE & FALSE & FALSE & FALSE & FALSE \\
        nomad2018-predict-transparent-conductors & 0.09346 & 0.05589 & 0.06229 & 0.06582 & 0.06988 & FALSE & FALSE & FALSE & FALSE & FALSE & TRUE  \\
        plant-pathology-2020-fgvc7              & 0.74045 & 0.97836 & 0.97465 & 0.97361 & 0.94852 & FALSE & FALSE & FALSE & FALSE & FALSE & TRUE  \\
        random-acts-of-pizza                    & 0.64037 & 0.97908 & 0.76482 & 0.69210 & 0.59960 & FALSE & FALSE & FALSE & FALSE & TRUE  & TRUE  \\
        ranzcr-clip-catheter-line-classification & -      & 0.97357 & 0.97152 & 0.97090 & 0.96750 & FALSE & FALSE & FALSE & FALSE & FALSE & FALSE \\
        spooky-author-identification            & 1.08483 & 0.16506 & 0.26996 & 0.29381 & 0.41879 & FALSE & FALSE & FALSE & FALSE & FALSE & TRUE  \\
        tabular-playground-series-dec-2021      & 0.95746 & 0.95660 & 0.95658 & 0.95658 & 0.95342 & TRUE  & TRUE  & FALSE & FALSE & TRUE  & TRUE  \\
        tabular-playground-series-may-2022      & -       & 0.99823 & 0.99822 & 0.99818 & 0.97268 & FALSE & FALSE & FALSE & FALSE & FALSE & FALSE \\
        text-normalization-challenge-english-language & -   & 0.99012 & 0.98232 & 0.97592 & 0.97591 & FALSE & FALSE & FALSE & FALSE & FALSE & FALSE \\
        tgs-salt-identification-challenge       & 0.06620 & 0.89097 & 0.87163 & 0.85918 & 0.81665 & FALSE & FALSE & FALSE & FALSE & FALSE & TRUE  \\
        \bottomrule
    \end{tabular}
    }
    \label{tab:sweagent_claude_results}
\end{table}


\end{document}